\documentclass[fleqn,usenatbib]{aa}

\usepackage{newtxtext,newtxmath}

\usepackage[T1]{fontenc}
\usepackage{ae,aecompl}

\usepackage{graphicx}	
\usepackage{amsmath}	
\usepackage{lipsum}
\usepackage{rotating}
\usepackage{pdflscape}
\usepackage[usenames,dvipsnames,svgnames,x11names]{xcolor}
\usepackage{ulem}
\usepackage[colorlinks=true, linktocpage, linkcolor={blue!60!black}, citecolor={blue!60!black}, urlcolor={blue!60!black}]{hyperref}
\usepackage{pdfpages}
\usepackage{multirow}
\usepackage{booktabs}
\usepackage{xspace}

\def\cluster{MACS~J0717.5+3745\xspace}
\def\macscluster{MACS~J0717.5+3745\xspace}
\def\sz{Sunyaev-Zeldovich\xspace}

\def\chandra{{Chandra}\xspace}
\def\rosat{{ROSAT}\xspace}
\def\rass{{ROSAT All Sky Survey}\xspace}
\def\xmm{{XMM-Newton}\xspace}

\def\nustar{{NuSTAR}\xspace}
\def\xrism{{XRISM}\xspace}
\def\athena{{NewAthena}\xspace}

\def\mustang{{MUSTANG2}\xspace}

\begin{document}




\title{The thermodynamic structure and large-scale structure filament in \cluster{}}

\author{J. P. Breuer\inst{1,2}\thanks{E-mail: jeanpaul.breuer@gmail.com}
\and
N. Werner\inst{1}
\and
T. Pl\v{s}ek\inst{1}
\and
F. Mernier\inst{3,4,5}
\and
K. Umetsu\inst{6}
\and
A.~Simionescu\inst{7,8,9}
\and
M.~Devlin\inst{10}
\and
L.~Di~Mascolo\inst{11,12}
\and
T.~Dibblee-Barkman\inst{13}
\and
S.~Dicker\inst{10}
\and
B.~S.~Mason\inst{14}
\and
T.~Mroczkowski\inst{15}
\and
C.~Romero\inst{16}
\and
C.~L.~Sarazin\inst{17}
\and
J.~Sievers\inst{13}
}

\institute{Department of Theoretical Physics and Astrophysics, Masaryk University. Kotl{\'a}{\v r}sk{\'a} 2, Brno, 611 37, Czech Republic
\and
Department of Physics, Graduate School of Advanced Science and Engineering, Hiroshima University Kagamiyama, 1-3-1 Higashi-Hiroshima, 739-8526, Japan
\and
NASA Goddard Space Flight Center, Code 662, Greenbelt, MD 20771, USA
\and
Department of Astronomy, University of Maryland, College Park, MD 20742-2421, USA
\and
ESA/ESTEC, Keplerlaan 1, 2201 AZ Noordwijk, The Netherlands
\and
Academia Sinica Institute of Astronomy and Astrophysics (ASIAA), No. 1, Section 4, Roosevelt Road, Taipei 106216, Taiwan
\and
SRON Netherlands Institute for Space Research, Niels Bohrweg 4, 2333CA Leiden, The Netherlands 
\and
Leiden Observatory, Leiden University, PO Box 9513, 2300 RA Leiden, The Netherlands
\and
Kavli Institute for the Physics and Mathematics of the Universe (WPI), The University of Tokyo, Kashiwa, Chiba 277-8583, Japan
\and
{University of Pennsylvania, 209 S. 33rd St., Philadelphia, PA 19014, USA}
\and
Kapteyn Astronomical Institute, University of Groningen, Landleven 12, 9747 AD, Groningen, The Netherlands
\and
Laboratoire Lagrange, Université Côte d’Azur, Observatoire de la Côte d’Azur, CNRS, Blvd de l’Observatoire, CS 34229, 06304 Nice cedex 4, France
\and
{Department of Physics, McGill University, 3600 University Street Montreal, QC, H3A 2T8, Canada}
\and
{National Radio Astronomy Observatory, 520 Edgemont
  Rd., Charlottesville VA 22903, USA} 
\and
European Southern Observatory, Karl-Schwarzschild-Stra{\ss}e 2, 85748 Garching bei M{\"u}nchen, Germany
\and
{Center for Astrophysics, Harvard \& Smithsonian, 60 Garden Street, Cambridge, MA 02138, USA}
\and
Department of Astronomy, University of Virginia, P.O. Box 400325, Charlottesville, VA 22904, USA
}

\date{Accepted XXX. Received YYY; in original form ZZZ}



\abstract{
  We present the results of \chandra{} and \xmm{} X-ray imaging and spatially resolved spectroscopy, as well as new \mustang{} 90~GHz observations of the thermal Sunyaev-Zeldovich effect from \macscluster{}, an intermediate redshift ($z=0.5458$) and exceptionally massive ($3.5\pm0.6\times10^{15}$ M$_\odot$) Frontier Fields cluster experiencing multiple mergers and hosting an apparent X-ray bright large-scale structure filament. Thermodynamical maps are produced from \chandra{}, \xmm{}, and \rosat{} data using a new method to model the astrophysical and instrumental backgrounds. The temperature peak of $24\pm4$ keV is also the pressure peak of the cluster and is spatially closely correlated with the \sz{} peak from the \mustang{} data. We characterize a potential shock candidate at the cluster center, based on the sharp temperature and pressure gradient, and quantify its temperature-derived Mach number in various directions to span a range of $\mathcal{M} = (1.7 - 2.0) \pm 0.3$. Bayesian X-ray Analysis methods were used to disentangle different projected spectral signatures for the filament structure, with Akaike and Bayes criteria being used to select the most appropriate model to describe the various temperature components. We report an X-ray filament temperature of $3.1_{-0.3}^{+0.6}$ keV and a density $(3.78\pm0.05)\times10^{-4}\,{\rm cm^{-3}}$, corresponding to an overdensity of $\sim400$ relative to the critical density of the Universe. We estimate the hot gas mass of the filament to be $\sim6.1\times10^{12}~\rm M_\odot$, while its total projected weak lensing measured mass is  $\sim(6.8\pm2.7)\times10^{13}~\rm M_\odot$, indicating a hot baryon fraction of 4--10\%.} 

{}

\maketitle



\section{Introduction}
Current cosmological models and numerical simulations predict that the majority of the missing baryons in our Universe sit in the faint galaxy cluster outskirts and the interconnecting filaments of the cosmic web \citep{baryons1, Dave2001}. This warm-hot intergalactic medium (WHIM) contributes to the growth of galaxy clusters through a slow, constant accretion, as the infalling filamentary gas virializes within their gravitational potential wells. 

Only a small number of X-ray bridges and filaments have been studied so far. This is partially because the diffuse WHIM remains a complicated challenge to detect with today's instruments. Numerical cosmological simulations predict WHIM temperatures of $10^5$ to $10^7$ K and densities of 10$^{-7}$ to 10$^{-4}$ cm$^{-3}$ \citep{baryons2, cosmicweb2016}; in other words, features in the WHIM are inherently characterized by soft X-ray emission, low surface brightness, and poor signal to noise ratios due to dominant contributions to the signal from both astrophysical and instrumental backgrounds. Needless to say, much care needs to be taken when modelling the spectra of these complex regions to account for all of the different contributions from other emission components. 

Current observational information about filaments is derived from either absorption or emission. Observations in emission have so far been limited to a small number of systems, \macscluster{} \citep[][]{Filament2004}, Abell 399/401 \citep[][]{A399-401}, Abell 222/223 \citep[][]{A222-223}, Abell 2811 (+offset)/2804/2801 \citep[][]{A2811-2804-2801}, Abell 3558/3556 \citep[][]{A3558-3556}, Abell 2744 \citep[][]{A2744}, Abell 3391/3395 \citep[][]{A3395-3391,A3391-95a,A3391-95b,A3391-95c}, Abell 2029/2033 \citep[][]{A2029-2033}, Abell 98N/98S \citep[][]{premerger_shock}, Abell 3667/3651 \citep[][]{A3667-3651}, and A3530/32 and A3528-N/S \citep[][]{Migkas2025}. Temperature measurements were obtained only for five of these systems. For two of the five, it has been reported that the filament emission is dominated by the emission of the Intracluster Medium (ICM), as evidenced by the excessively large temperatures.

There is a debate about the origin of these X-ray filaments. While some of the detected bridges could indeed be classified as WHIM filaments, others are due to other physical processes, such as shock-related compression heating between the two merging atmospheres or emission of ram-pressure stripped tails from infalling haloes. 

One of the most extensively studied merging galaxy clusters in almost every available wavelength is \macscluster{} (RA 07h17m32.1, DEC +37$^\circ$45'21"), an extremely massive, intermediate redshift ($z = 0.5458$), Hubble Frontier Fields Cluster \citep{ebelling_ffields}. \macscluster{} is also one of the most complex galaxy clusters known to date, having hosted at least four different sub-cluster collisions. The collective mass from the many ingested dark matter halos makes this a great target for gravitational lensing studies and additional exploration of local substructures via the \sz{} effect due to the high temperatures. The mergers have also resulted in a complex ICM morphology, characterized by discontinuities, such as shock fronts and cold fronts, as well as many peculiar features in the X-ray and radio bands. Besides the gravitational lensing study, \citet{Jauzac2018}, also looked at many of these X-ray substructures in detail. Most intriguing is the prominent X-ray bridge located in the S-SE of the cluster, which has previously been studied by \citet{VanWeeren2016,VanWeeren2017} using \chandra{} data.

This paper aims to investigate the X-ray bridge in detail, statistically disentangling the diverse distribution of complex emission signatures associated with the filament using Bayesian X-ray Analysis (BXA) methods and nested models, along with a new dynamically-adaptive method for the full instrumental background modelling of each region for both \chandra{} and \xmm{}. This paper also presents new thermodynamical maps using a joint modelling method combining available data from \chandra{}, \xmm{}, and \rosat{}, as well as trend-modeled residual maps where the average cluster thermodynamical properties over radial distances are removed. This paper additionally presents previously unpublished data from 90~GHz \mustang{} observations of the \sz{} effect in the direction of \macscluster{}. In Section~\ref{sec:observations}, we explain the methods used in the data reduction, followed by Sec.~\ref{sec:background} that explains the instrumental and X-ray background model and analysis of both \chandra{} and \xmm{}. Section~\ref{sec:results} then describes the main results of the thermodynamic maps and the properties of the filament, while Section~\ref{sec:discussion} discusses these results. Throughout the paper, we assume the standard $\Lambda$ cold dark matter cosmology with $\Omega_m = 0.286$, $\Omega_\Lambda = 0.714$, and $H_0 = 69.6$ \citep{StandardLambdaCosmo}. Consequently, at $z = 0.5458$, 1 arcmin corresponds to 387.42 kpc. The wilms abundance table is adopted for all plasma emission and photoelectric absorption models in the discussed spectral models \citep{wilms}. Unless stated otherwise, the error bars correspond to a 68\% confidence interval.

\section{Observations and Data Reduction}\label{sec:observations}

\subsection{Chandra Observations}\label{sec:chandra}
The data from four Chandra pointings (see Table~\ref{tab:observations}) were reprocessed using \texttt{repro} from the level-1 event lists with the standard software packages using the most recent versions of CIAO and CALDB (versions 4.16.0 and 4.11.0). The good time intervals (GTI) of the observations after filtering periods of flaring are summarized in Table~\ref{tab:observations}. All observations were used for creating the surface brightness images; however, only observations 4200, 16235, and 16305 were considered in further spectral analysis, as observation 1655 experienced residual soft proton flaring contamination which would bias spectral analysis. Totalling all available Chandra observations gives approximately 228.9 ks of observing time; however, with the removal of observation 1655, only 211.8 ks were used for spectral analysis. Because of the low signal-to-noise ratio, blank-sky background files were explicitly not used in favour of using epoch-specific stowed background files normalized by the data over background count ratio in the 10-12 keV energy range. When dealing with extremely faint X-ray substructures, the stowed background files give additional control over the systematics in the background. Merged broadband images were produced in the 0.5 to 7 keV band, shown in Fig.~\ref{fig:fovboth} \textit{(left)}. 

\subsection{XMM-Newton Observations}\label{sec:xmm}
Complementary to the \chandra{} data, we also retrieved archival \xmm{} observations of \macscluster{}, as listed in Table~\ref{tab:observations}. Three pointings showing sufficient data quality were downloaded and reduced using both the XMM Science Analysis System (SAS) software (v20.0.0) and the current calibration files (CCF) by following the procedure detailed in the Extended Source Analysis Software (ESAS v20.0.0) cookbook. To get cleaned event files for each detector, we used the standard \texttt{emchain} and \texttt{mos-filter} routines for the MOS data, and the \texttt{epchain} and \texttt{pn-filter} routines for the pn data. \texttt{mos-filter} and \texttt{pn-filter} remove periods of anomalously high count rate by calculating GTIs through the \texttt{espfilt} routine. The total GTI for each observation are also shown in Table~\ref{tab:observations}.

\begin{figure*}
	\includegraphics[width=\textwidth]{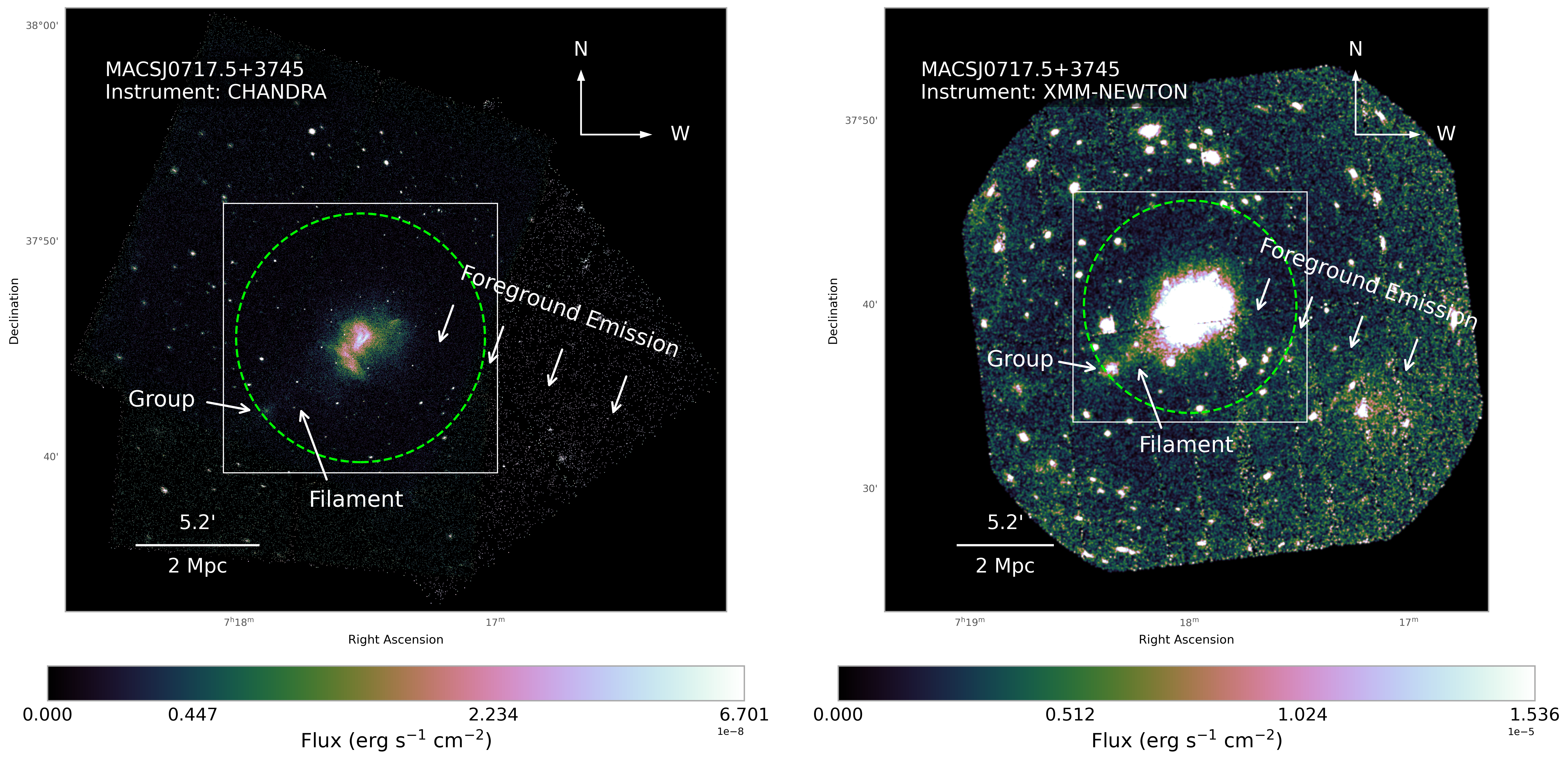}%
	\caption{Images of the full FOV of \chandra{} (left) and \xmm{} (right) of \macscluster{}. The surface brightness excess to the west/south-west are several foreground structures in between us and the cluster, indicated by the `Foreground Emission' label. The \xmm{} image is over-saturated in the cluster center so that this can be seen more clearly. The green dashed circle indicates the approximate $R_{200}$ of the cluster. Intensity scale is in flux per pixel (0.492$\times$0.492 arcsec for Chandra, and 4.1$\times$4.1 arcsec for XMM pn).}
	\label{fig:fovboth}
\end{figure*}

\begin{figure*}
	\includegraphics[width=\textwidth]{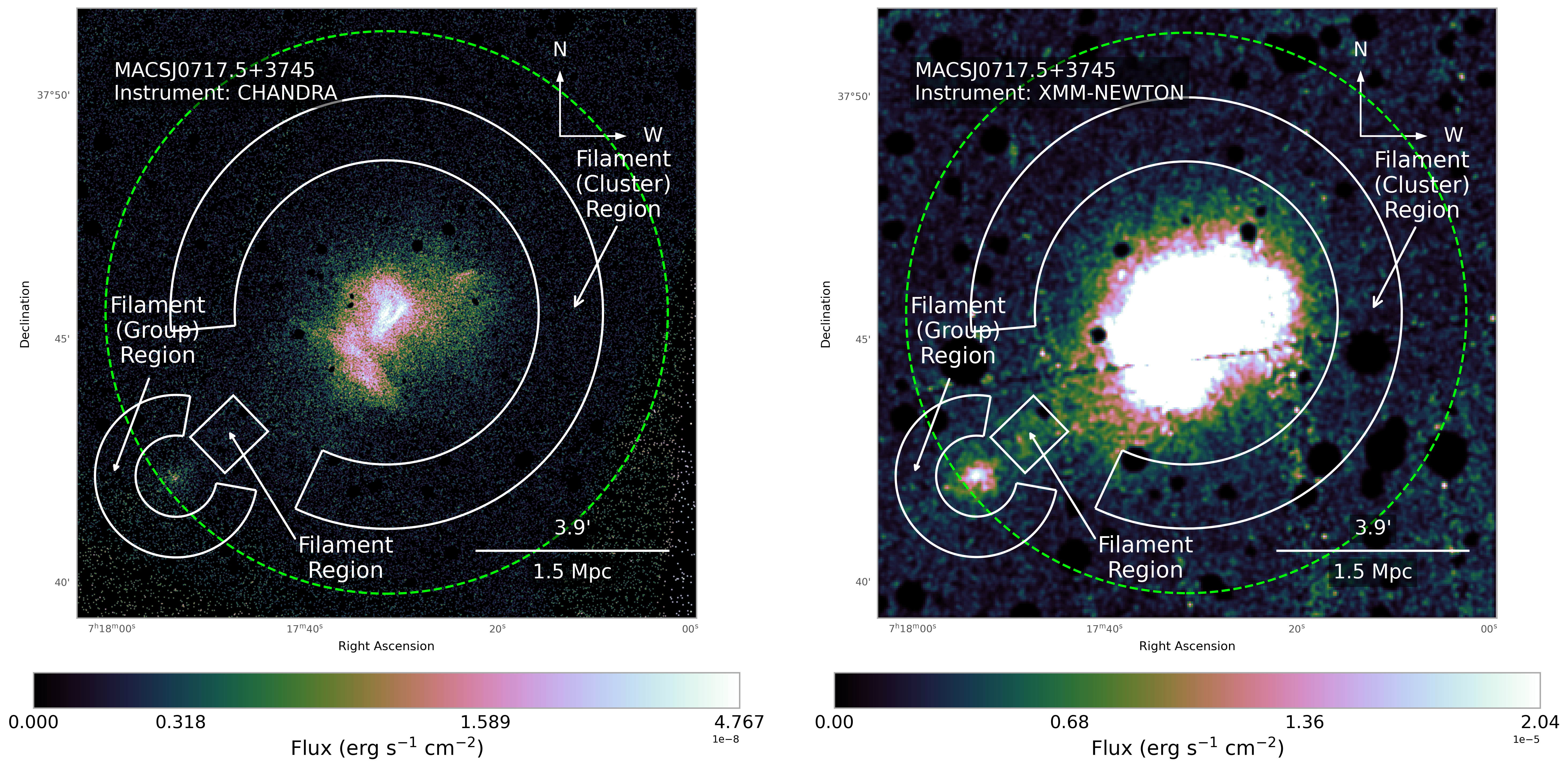}%
	\caption{The \chandra{} (left) and \xmm{} (right) images of \macscluster{}, restricted to the same field of view as the \sz{} data and thermodynamical maps, visible as the box in Fig.~\ref{fig:fovboth}. Intensity scale is in flux per pixel (0.492$\times$0.492 arcsec for Chandra, and 4.1$\times$4.1 arcsec for XMM pn). Part of the \xmm{} image is over-saturated so that the filament structure is more visible. The white lines indicate the regions used for modelling the spectra of the outskirts of the cluster and the group together with the filament. The green dashed circle indicates the approximate $R_{200}$ of the cluster.}
\label{fig:squareboth}
\end{figure*}

Images were extracted for each detector of each observation using the ESAS tasks \texttt{mos-spectra} and \texttt{pn-spectra}, while background images were created using the ESAS tasks \texttt{mos\_back} and \texttt{pn\_back}, before being combined using the \texttt{comb} routine. Point sources were detected using a wavelet detection algorithm, \texttt{wavdetect}, by first making a psfmap for \xmm{} with a constant size of 9 arcsec. This is based on the estimate that at 1.5 keV, the 1-$\sigma$ integrated volume of a 2D Gaussian for each pixel ranges from 7 arcseconds on-axis to around 11 arcseconds at edge of the FOV, and our selected box region of interest is near the central pointing. Merged broadband images from all observations were produced in the 0.5 to 7 keV band using the \texttt{merge\_comp\_xmm} routine, shown in Fig.~\ref{fig:fovboth} \textit{(right)}. A zoomed-in, over-saturated view of the cluster center and filamentary structure (with pointsources removed) can be seen in Fig.~\ref{fig:squareboth}.

The spectra of all regions are extracted from each \xmm{} observation using the standard ESAS tools \texttt{mos-spectra}, \texttt{mos\_back}, \texttt{pn-spectra} and \texttt{pn\_back}. The \xmm{} regions are first converted into observation and instrument-specific detector coordinates; however, because of the many point sources and complex region geometries, the complete region expression becomes too long for the fits data subspace. Therefore it is necessary to first turn each sky region minus any overlapping point sources into a region file in detector coordinates for extraction with ESAS. 

All spectra from both \chandra{} and \xmm{} were optimally binned using the method described in \citet{optbin}, and all astrophysical parameters were linked through all spectra for a joint fit. Unless otherwise specified, all spectra in this work were produced in a similar way. These regions were finally fit using \texttt{SHERPA} version 4.16.0 (from \texttt{CIAO-4.16}), using the \texttt{XSPEC} version 12.13.1e model library and AtomDB version 3.0.9. 

\begin{table}
	\centering
	\caption{Summary of the data, excluding overly contaminated observations. GTI is the net exposure good time intervals for all instruments after cleaning the data. $(*)$ Not used.}
	\label{tab:observations}
	\begin{tabular}{lccr}
	  \hline 
		Telescope & ObsID & Date & GTI (ks)\\
        \hline 
        \textit{Chandra} & 1655$^{(*)}$ & 2001-01-29 & 17.1\\		
        \textit{Chandra} & 4200 & 2003-01-08 & 57.2\\
	\textit{Chandra} & 16235 & 2013-12-13 & 67.9\\		
	\textit{Chandra} & 16305 & 2013-12-11 & 86.7\\
        \textit{XMM-Newton} & 0672420101 & 2011-10-11 & 41.7\\
        \textit{XMM-Newton} & 0672420201 & 2011-10-13 & 51.3\\        
        \textit{XMM-Newton} & 0672420301 & 2011-10-15 & 43.4\\
        \hline 
	\end{tabular}
\end{table}

\subsection{CLASH Weak Lensing Observations}
In this study, we use ground-based weak-lensing data products from the CLASH program, as presented in \citet{clash}. These products are derived from deep $BVR_\mathrm{C}i'z'$ imaging with Subaru/Suprime-Cam, complemented by UV imaging from Megaprime/MegaCam and near-IR observations from WIRCAM on the Canada--France--Hawaii Telescope \citep[see also][]{medezinski_clash}. For further details on data reduction and weak-lensing analysis, refer to \citet{clash}.

\subsection{\mustang{} \sz{} Observations} \label{sec:sz_observations}
\cluster{} was observed by the 100-meter Green Bank Telescope (GBT) under project codes AGBT17A\_340 and AGBT17B\_266 (in semesters 2017A \& 2017B, respectively) using the MUSTANG-2 instrument, a 215-element continuum bolometer array operating at 90~GHz \citep{Dicker2014}. 
The observations were conducted using Lissajous daisy scans with radii of 3\arcmin\ and 3\farcm5 \citep{Romero2020, Romero2023}.  
The data processing was identical to that in, for example, \cite{Romero2020}.  The resulting map, produced using the MUSTANG-2 pipeline, is shown in the lower left panel of Fig. \ref{fig:sz}. The total integration time on target was $2.6$~hours and the resulting map has a resolution of $10''$, with an RMS noise level of $\sim 28 \, {\rm \mu Jy/bm}$. Besides the astrophysical contaminants and the instrumental noise, \mustang{} data is affected by large-scale filtering of the astrophysical signal, which affects the imaging quality.

\section{Background analysis}\label{sec:background}
\subsection{Chandra Stowed Instrument Background}\label{sec:chandra_background}

The \chandra{} Stowed Background files are generated in a similar procedure as the blank-sky background files and were scaled by the data over background count ratio in the 10-12 keV energy range. The \chandra{} instrumental background can be described as a mixture of various signal contributions across different energy bands. Specifically, cosmic rays hit the detectors from every direction and create a continuum emission. Additionally, the particle interactions can produce instrumental X-ray fluorescence lines from the various elements in and around the detector.

Of the available background datasets, the total stowed background merged exposure time was 2.91 Ms for the year 2000, 2.57 Ms for 2005, and 1.68 Ms for 2009. Spectra were extracted from the full field of view of each available year and were modelled following a similar process as described in \citet{chandra_bkg_description} for \chandra{} and \citet{Mernier2015} for \xmm{}, with independent fits for each of the detectors. Special care was also taken regarding the Faint or Very Faint imaging modes, resulting in modified models that involve these different cases. The stowed background data was modelled in the 0.7 keV to 10 keV band for both the ACIS-I and ACIS-S detectors. The continuum emission is defined by a broken power-law for the Faint imaging submode and by a power-law in the Very Faint imaging submode, while all instrumental lines were modelled using Gaussians and frozen to their best fits. \citet{acis_shape} noted that the shape of the particle background is fairly consistent, and only the normalizations in different epochs are changing from year to year; therefore, we refrain from including redundant information for all years, and give detailed results of the continuum emission and instrumental lines in Table~\ref{tab:chan_combined_table}. Elements corresponding to line energies used in our background model were documented in other work \citep{1967_xray_table, suzuki21}.

The line energies which did not exactly correspond to reported lines were marked in the table with `$*$', while the 2.6 keV line was marked with `$**$' as it was not previously mentioned in other reference material. This line in particular showed up only in the ACIS-I stowed backgrounds and is unlikely to be an instrumental line, rather some artifact of the response itself. It was included in the ACIS-I model as an empirical improvement for completeness. All \chandra{} spectral fits reported in this paper were performed in the 0.7 to 7.0 keV range.

\setlength{\tabcolsep}{1.4pt}
\renewcommand{\arraystretch}{1.4}
\begin{table}
\centering
\caption{Summary of the ACIS-I and ACIS-S Stowed Background Data: Best-fit parameters of the continuum emission components and summary of instrumental lines, used as initial parameters for the automatic background fitting routine. Line energies with $^{*}$ indicate uncertainty in the Element/Type, represented by closest reported lines, $^{**}$ indicates an undocumented line. These parameters translate to \textsc{Xspec} models as \textit{powerlaw} or \textit{bknpower} with the additive gaussians \textit{gauss}.}
\label{tab:chan_combined_table}
\begin{tabular}{lllll}
\toprule
\multicolumn{5}{c}{Best-fit Parameters of the Continuum Components} \\
\midrule
& \multicolumn{2}{l}{Faint (bknpower)} & \multicolumn{2}{l}{Very-Faint (powerlaw)} \\
Parameters & ACIS-I & ACIS-S & ACIS-I & ACIS-S \\
\midrule
$\Gamma$ & 2.386 & 2.386 & 0.349 & 0.349 \\
$E_{\text{break}}$ (keV) & 0.692 & 0.692 & - & - \\
$\Delta \Gamma$ & 0.418 & 0.418 & - & - \\
$\text{norm}$ & 0.104 & 0.104 & 0.194 & 0.194 \\
\addlinespace 
\midrule
\multicolumn{5}{c}{Summary of Instrumental Lines} \\
\midrule
\multicolumn{2}{c}{ACIS-I} & \multicolumn{2}{c}{ACIS-S} & \\
Energy (keV) & Element & Energy (keV) & Element & \\
\midrule
0.529 & - & - & - & \\
1.490 & Al K$\alpha$ & 1.490 & Al K$\alpha$ & \\
1.786 & Si K$\alpha$ & 1.786 & Si K$\alpha$ & \\
2.141$^{*}$ & Au M$\alpha_1$/M$\alpha_2$ & 2.142$^{*}$ & Au M$\alpha_1$/M$\alpha_2$ & \\
2.6$^{**}$ & - & - & - & \\ 
7.478 & Ni K$\alpha$ & 7.478 & Ni K$\alpha$ & \\
8.3$^{*}$ & Ni K$\beta$/Au L$_1$ & - & - & \\
9.74$^{*}$ & Au L$\alpha_1$/L$\alpha_2$ & 9.740$^{*}$ & Au L$\alpha_1$/L$\alpha_2$ & \\
- & - & 9.755$^{*}$ & Au L$\alpha_1$/L$\alpha_2$ & \\
\bottomrule
\end{tabular}
\end{table}

\subsection{XMM-Newton Instrumental Background}\label{sec:xmm_bkg}
Similar to the \chandra{} Stowed Backgrounds, the \xmm{} Filter Wheel Closed instrumental backgrounds can also be described as a mixture of various spectral contributions across different energy bands. Besides the `hard particle' (HP), or quiescent particle background (QPB) components shared with \chandra{}, \xmm{} cluster source observations can additionally suffer from a residual contribution from a `soft-particle' (SP) component; remnants of the quiescent soft proton contamination, which may still pervade throughout the data despite having been excised from the observation during the GTI cleaning step. These `soft-protons' are particles trapped in Earth's magnetosphere, which are accumulated and concentrated by the flight optics during the orbit cycle, affecting around 40\% of all observation time.

The hard particle components were modelled using all currently available FWC data, corresponding to a total merged exposure time of around 4.77 Ms. Specifically, around 1.958 Ms of data from MOS1, 1.924 Ms of data from MOS2, and 0.8841 Ms of Extended Full-Frame pn data. Spectra were extracted from the full field of view of all available FWC observations, and modelled following a similar process as described in \citet{Mernier2015}, also with independent fits for each of the detectors. 

The FWC data were modelled in the 0.3 keV to 10 keV band for the MOS1 and MOS2 detectors and in the 0.4 keV to 10 keV range for pn. The continuum emission is fitted by a broken power-law, while all elemental lines were modelled using Gaussians described in Table~\ref{tab:xmm_combined_table}. Because these HP components are noise contributions directly from the detectors and are unrelated to the mirrors, these instrumental background models are not convolved by the Ancillary Response File (ARF).

The residual SP components can be also described as an ARF-unfolded power-law component with an index limited to a range between 0.1 and 1.4, which affects source spectra. However, the inclusion of the additional power-law did not contribute to an improvement in the background model fit statistics, indicating that the soft protons were fairly well treated in the data reduction.

To further check for any residual soft proton contamination, we compared the area-corrected count rates between the 'inside' and 'outside' regions of the Field of View (FOV) for each detector in each observation, shown in Table~\ref{tab:finfout} \citep{deluca_molendi2004}. All of the diagnostic values are under the lowest soft proton threshold of 1.15, indicating that none of the event lists are additionally contaminated by soft protons. As such, in this paper, the soft proton parameters were removed from the background model, which reduced the overall complexity. All \xmm{} spectral fits reported in this paper were performed in the 0.45 to 7.0 keV range for MOS and 0.3 to 7.0 keV for pn.

\setlength{\tabcolsep}{6.4pt}
\renewcommand{\arraystretch}{1.4}
\begin{table}
\centering
\caption{Comprehensive summary of MOS1, MOS2, and pn Filter Wheel Closed Data including Hard Particle components and Instrumental Lines used as initial parameters for the automatic background fitting routine. $(M1)$ or $(M2)$ indicate EPIC MOS1 or MOS2 only.  These parameters translate to \textsc{Xspec} models as \textit{powerlaw} or \textit{bknpower} with the additive gaussians \textit{gauss}.}
\label{tab:xmm_combined_table}
\begin{tabular}{lllll}
\toprule
\multicolumn{5}{c}{Best-fit Parameters of the Hard Particle Components} \\
\midrule
Parameters & MOS1 & MOS2 & pn & \\
\midrule
$\Gamma$ & 2.30 & 2.42 & 6.05 & \\
$E_{\text{break}}$ (keV) & 0.58 & 0.52 & 0.48 & \\
$\Delta \Gamma$ & 0.24 & 0.26 & 0.34 & \\
$\text{norm}$ & 4.02 & 4.34 & 0.12 & \\
\midrule
\multicolumn{5}{c}{Summary of Instrumental Lines} \\
\midrule
\multicolumn{2}{c}{MOS} & \multicolumn{2}{c}{pn} & \\
Energy (keV) & Element & Energy (keV) & Element & \\
\midrule
- & - & 0.363 & - & \\
0.521$^{(M1)}$ & - & - & - & \\
0.62 & - & - & - & \\
0.750 & - & - & - & \\
0.942$^{(M2)}$ & - & 0.958 & - & \\
1.489 & Al K$\alpha$ & 1.479 & Al K$\alpha$ & \\
1.746 & Si K$\alpha$ & - & - & \\
2.146 & Au M$\alpha$ & 2.100 & Au M$\alpha$ & \\
5.410 & Cr K$\alpha$ & 4.520 & Ti K$\alpha$ & \\
5.900 & Mn K$\alpha$ & 5.410 & Cr K$\alpha$ & \\
6.400 & Fe K$\alpha$ & 6.375 & Fe K$\alpha$ & \\
7.100 & Fe K & 7.450 & Ni K$\alpha$ & \\
7.480 & Ni K$\alpha$ & 8.011 & Cu K$\alpha$ & \\
8.100 & Cu K$\alpha$ & 8.580 & Zn K$\alpha$ & \\
8.640 & Zn K$\alpha$ & 8.860 & Cu K$\beta$ & \\
9.710 & Au L$\alpha$ & 9.520 & Zn K$\beta$ & \\
\bottomrule
\end{tabular}
\end{table}

\begin{table}
	\centering
	\caption{Summary of the \xmm{} $F_{in}/F_{out}$ data, indicating the soft proton contamination ratio between the inside and outside the FOV of the CCDs. Based on the ratios, none of the detectors on any observation were contaminated by soft protons.}
        \label{tab:finfout}
	\begin{tabular}{lccc}
	  \hline
		ObsID & $F_{in}/F_{out}$ M1 & $F_{in}/F_{out}$ M2 & $F_{in}/F_{out}$ pn \\
        \hline
        0672420101 & $1.102\pm0.027$ & $1.036\pm0.023$ & $1.027\pm0.023$\\
        0672420201 & $1.032\pm0.024$ & $1.037\pm0.022$ & $1.015\pm0.021$\\        
        0672420301 & $1.050\pm0.026$ & $1.035\pm0.024$ & $1.026\pm0.023$\\
        \hline 
	\end{tabular}
\end{table}

\subsection{The Astrophysical X-ray Background}
With the instrumental backgrounds accounted for, special attention must be given to carefully model the astrophysical components of the background before we can accurately constrain any of the physical parameters of the cluster component of the signal. Especially in the faint signal-to-noise (S/N) and signal-to-background (S/b) regimes, such as cluster outskirts or filamentary structures, the parameters are all very sensitive, so any small changes to the instrumental background can affect the astrophysical background, which can both collectively bias the final results.

The Cosmic X-ray Background (CXB) includes the unresolved point sources, distant Active Galactic Nuclei (AGN), typically modelled with a photoelectrically absorbed power-law with slope frozen at 1.41 and a free normalization \citep{cxb_powerlaw_markevitch}. The Galactic foreground sources can be described as a combination of several plasmas in collisional ionisation equilibrium (\texttt{apec}). They include the Local Hot Bubble (LHB), the Galactic Halo (GH), and recently another component, interpreted as a super-virialized (SV) component of our Galactic Halo, sometimes referred to as the `Galactic Corona'. The LHB is an approximately 0.1 keV plasma originating from a local supernova remnant that we are currently residing in. It can be effectively modelled using the \rosat{} \textit{All Sky Survey} (RASS) data with the support of the soft X-ray bands from \chandra{} and \xmm{}. Additionally, our Galactic Halo and its supervirial component are described by photoelectrically absorbed (\texttt{tbabs}, wilms), thermal \texttt{APEC} plasma models \citep{wilms}. For all of these local, foreground CXB components, the redshift was frozen to 0, the abundances frozen to solar, with the temperatures and normalizations free to fit. Unless otherwise stated, the 1.41 power law photon index was kept frozen. Meanwhile, the neutral hydrogen column density for the CXB and in all subsequent fits was frozen to $N_\text{H}$ = $8.36 \times 10^{20}$ cm$^{-2}$, corresponding to the sum of the atomic and molecular hydrogen column densities along the line of sight\footnote{https://www.swift.ac.uk/analysis/nhtot/} \citep{Kalberla05, galactic_abs}.

\begin{figure*}
        \includegraphics[width=\textwidth]{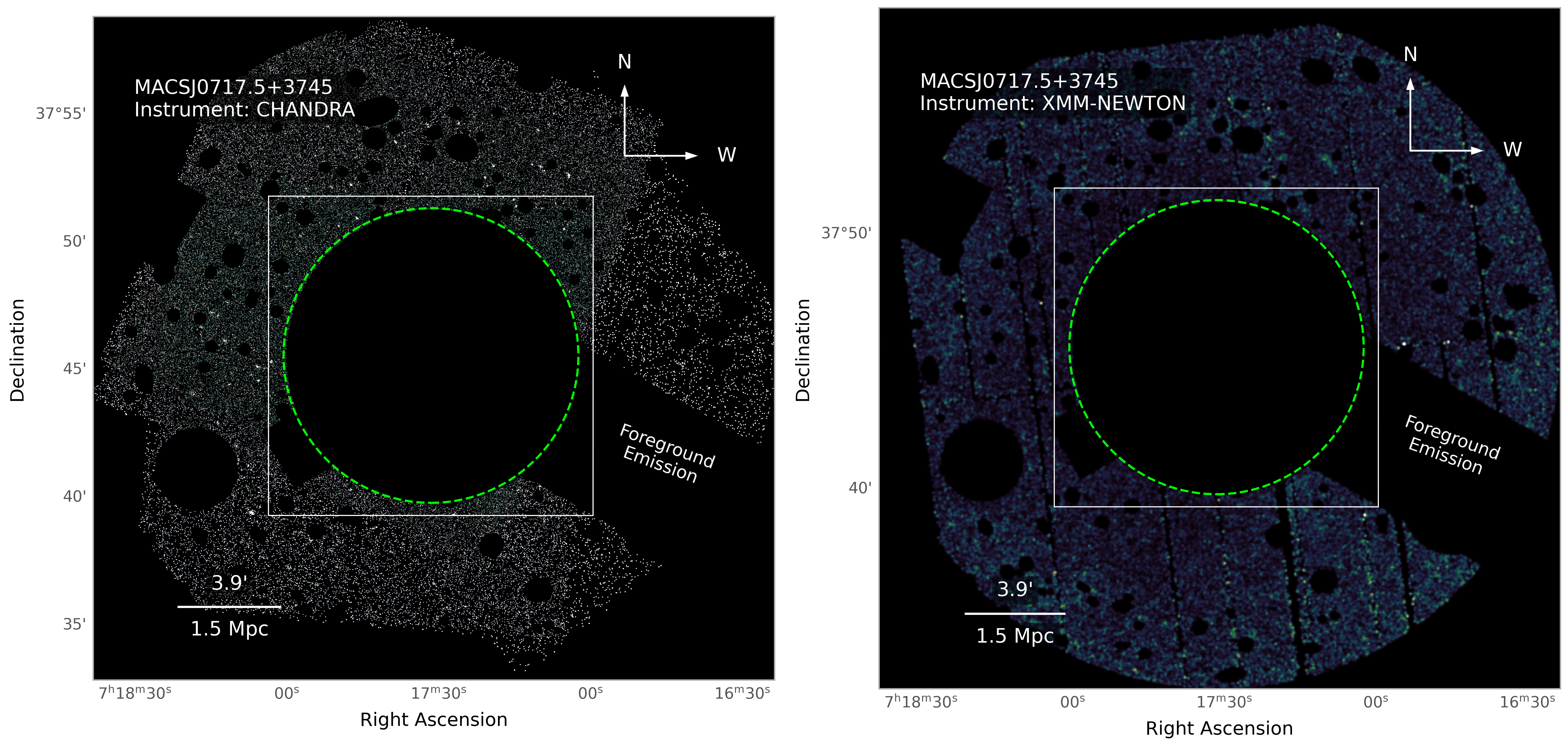}
	\caption{The emission-free regions taken from the observations to fit the CXB components, with \chandra{} on the left, and \xmm{} on the right. The box is the field of view used for Fig.~\ref{fig:squareboth} and the thermodynamical maps, corresponding to the total field of view of the Sunyaev Zeldovich data. The green dashed circle indicates the approximate $R_{200}$ of the cluster. The excluded circles are removed sources, the large excluded box region to the west corresponds to removed foreground galaxies, and another smaller box removes the emission of the filamentary structure.}
 \label{fig:cxbfov}
\end{figure*}

To constrain the local CXB parameters, we use a carefully constructed `blank field' for our observations by using all available data outside of the cluster $R_{200}$, from both the \chandra{} and \xmm{} data, as well as spectra from the \rass{}. The \rass{} spectrum was extracted from an annulus with radii 0.15 and 1 degrees around the cluster center, similar to \citet{VanWeeren2017}. We use a fit range for \rosat{} PSPC-C/B between 0.09 and 2.0 keV based on previous calibration database plots of on-axis effective area curves. Meanwhile, the \chandra{} and \xmm{} observations utilized the entire FOV of the observations with most point sources and emission regions removed. Additionally, the \chandra{} data excluded bright edges and the \xmm{} data also had approximately 1 arcminute from the edges of the observations removed, as can be seen in Fig.~\ref{fig:cxbfov}.

Using the available data from each instrument, the CXB was constrained using a total of 13 spectral data groups. The best-fit parameters of the foreground components are often degenerate with each other. For this reason the LHB, GH, SV, and unresolved point sources were determined using generous priors with the Bayesian X-ray Analysis tool, which uses a nested sampling algorithm (UltraNest\footnote{https://johannesbuchner.github.io/UltraNest/}) for Bayesian parameter estimation and model comparison, and is employed as a tool for faint S/N data within the AGN community \citep{ultranest,bxa}.

The top part of Table~\ref{tab:best_fits} shows the results of the different CXB components, with and without the inclusion of the \rass{} data, using both a background subtracted and background modelled fits. Without the \rosat{} data, the LHB temperature was fixed to 0.09 keV and a free normalization. Using all available data and the background models, the best-fit results using all 13 spectra gave a LHB temperature of 0.108$\pm 0.02$ keV, GH temperature of 0.156$\pm 0.002$ keV, and SV temperature of $0.697\pm0.011$ keV, with a SV normalization around 10\% that of the GH, consistent with other studies.

In all of the spectral analysis in this paper, spectral data was never combined, but instead, all data was fit simultaneously. For each of the spectral data groups, a 20\% systematic error was applied in Sherpa/Xspec to account for any cross-calibration related discrepancies in measurements between the instruments. A more detailed discussion regarding the cross-calibration can be found in Appendix~\ref{sec:cross_calibration}.

Once the CXB parameters are constrained, it is possible to more accurately constrain the cluster emission and, consequently, to create thermodynamical maps and trend-averaged maps of \macscluster{} to better study and visualize the ICM physics and internal structures. Specifically, we used a modified version of the contour binning algorithm from \citet{Sanders06}, which we first translated into Python from C++, and have released as \texttt{pycontbin}\footnote{https://github.com/jpbreuer/contbin\_python}. This contour binning algorithm creates regions with a pre-selected S/N ratio while conserving the general contours of the data by grouping neighbouring pixels of similar surface brightness and yields statistically independent regions. An added benefit of this modified contour binning code is that it can additionally take a static value for a Point Spread Function (PSF) as a constraint and merge map bins that are smaller than the given value, ensuring bins larger than the instrument PSF. The contour binning maps were derived from an exposure corrected \xmm{} mosaic image, scaled to have the approximate number of counts per pixel as the sum of observations. The smallest bin in this map corresponds to a circle of around 10.6 arcseconds radius, around the size of the PSF of \xmm{}. Like the CXB region, all contour bin map regions were converted into polygons and were extracted with both \chandra{} and \xmm{}, with the idea that the \chandra{} resolution could be exploited in the joint fitting procedure as a correction to the fit results in the cases where the complex polygon region geometries occasionally fall under the size of the \xmm{} psf. This resulted in a joint fit between 25 different spectra for each spatial bin, 13 from the CXB region, and 12 spectra from each spatial bin.

We again use the model for a plasma in collisional ionisation equilibrium (\texttt{apec}) with photoelectric absorption (\texttt{tbabs}), but now with fixed values for the metallicity at 0.3 Solar and redshift at $0.5458$. We additionally scale each region by its area in arcminute$^2$ using a model constant, so that we can simultaneously fit the large cxb region with each cluster spectra. The CXB parameters are first frozen to their best-fit values before fits are performed to constrain the cluster temperatures \textit{kT} and normalizations \textit{norm}. Finally, the ICM metallicity is thawed to constrain its value for each region. This fitting procedure is repeated for every bin and for every instrument independently before being performed on the combined data from all observatories and again for the combined data from all available instruments. Unless otherwise mentioned, this spectral fitting methodology applies to any other spectral analysis considered for the rest of the paper. 

\section{Results}\label{sec:results}
\subsection{Thermodynamical Maps}\label{sec:cluster-thermomaps}

Exploiting this joint-fitting methodology using \chandra{} and \xmm{} with the full instrumental background modelling gave around 110 spatial bins for the electron density map, around an order of magnitude improvement in the map resolution compared to the around 12 spatial bins shown in the previous map reported in \citet{VanWeeren2017}, which only exploited the available \chandra{} data and the blank-sky fields. Despite the smaller bin sizes, all map bins shown in this paper are larger than the \xmm{} PSF, and our large reported temperature of $\geq 20$ keV is consistent with the previous reported temperature in \citet{VanWeeren2017}.

\begin{figure*}
  \includegraphics[width=\textwidth]{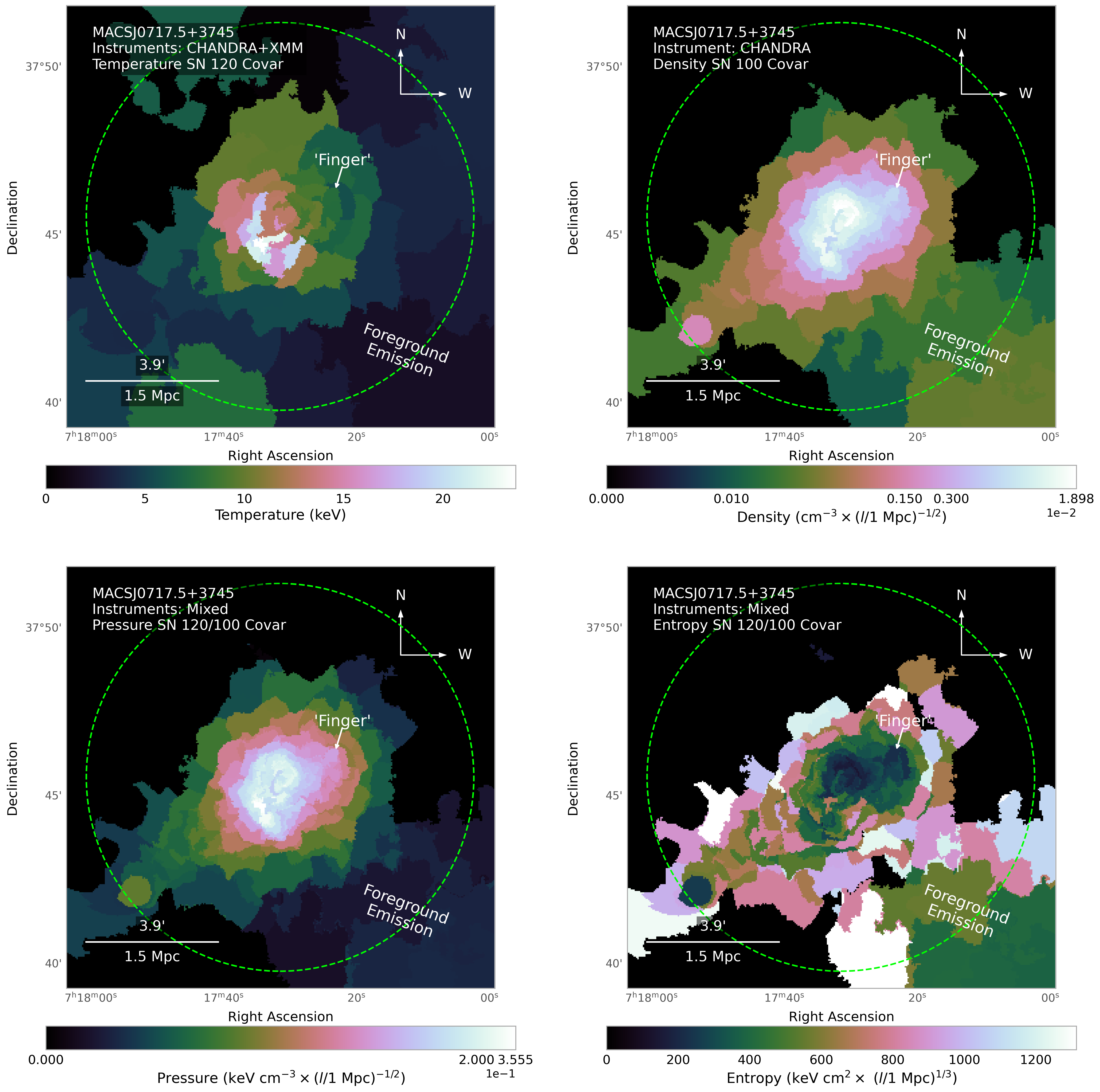}
  \caption{Maps of temperature (\textit{top left}), density (\textit{top right}), pressure (\textit{bottom left}), and entropy (\textit{bottom right}) with corresponding units of keV, cm$^{-3}\times$($l/1$Mpc)$^{-1/2}$, keVcm$^{-3}\times$($l/1$Mpc)$^{-1/2}$, and keVcm$^2\times $ ($l/1$ Mpc)$^{1/3}$. The temperature and electron density maps were created using binning with a S/N of 120 (14400 counts) and 100 (10000 counts), respectively, while the pressure and entropy maps were produced using the results from the top two S/N maps. Density maps were derived from the normalization fits assuming a 1 Mpc projected line of sight depth. Spectral fits were made using data from both \chandra{} and \xmm{} for the temperature, and only \chandra{} for the density, each using the CXB best fit results from \rosat{}, \chandra{}, and \xmm{}. Fits for these maps were performed using the most appropriate background model for each region according to the AIC. }\label{fig:1mpc}
\end{figure*}

Thermodynamical maps were computed from a scaled, exposure-corrected \xmm{} mosaic image, with S/N values of 120, 100, 70, and 50 (14400, 10000, 4900 and 2500 counts per bin, respectively). Regions from S/N 120 were extracted from \xmm{}, and the other bins from only \chandra{} to make use of the better spatial resolution. To compute asymmetric errors, we resample the multivariate distribution of the parameters from the covariance matrix provided by the fit results 100000 times, using the median and the 16th and 84th percentiles of the resampled distribution for the 68\% confidence intervals. Shown in Fig.~\ref{fig:1mpc} is the S/N 120 temperature map and S/N 100 density map, along with their combined S/N pressure and entropy maps, all of which had relative errors and propagated uncertainties lower than 20\%. Combining the S/N 120 and 100 region maps yields a better-resolved gradient between map regions at the expense of having completely statistically independent bins. This is our favoured method for improving the map resolution without suffering from over-smoothing, as seen in other binning methods. We discuss the maps in more detail in Sec.~\ref{sec:thermo_discussion}. 

Due to the nature of merging systems, projection effects may bias the results due to potential asymmetries along the projected line of sight. Moreover, because this cluster is a highly-disturbed, non-spherically symmetric, merging cluster with possibly more than 4 distinct sub-cluster populations in different merging states, it is difficult to easily approximate the integration volume due to uncertainties in the projected geometry. To begin with a simplification, we first assume a flat, pancake-like approximation for the line of sight depth of 1 Mpc for each map pixel in each bin, shown in Fig.~\ref{fig:1mpc}. The units given in the thermodynamic maps in Fig.~\ref{fig:1mpc} are, cm$^{-3}\times$($l/1$Mpc)$^{-1/2}$ for the density, keVcm$^{-3}\times$($l/1$Mpc)$^{-1/2}$ for the pressure, and keVcm$^2\times $ ($l/1$ Mpc)$^{1/3}$ for the entropy.

The temperature \textit{kT} and electron density \textit{$n_\textrm{e}$} were derived directly from the spectral fitting, although the electron density was determined from a normalization parameter, {\it norm}, formally defined as
\begin{equation}
\textit{norm} = \frac{10^{-14}}{4\pi (d_\textsc{a} (z+1))^2}\int n_\textrm{e} n_\textrm{p} dV \,,
\label{eq:norm}
\end{equation}
where $d_\textsc{a}$ is the angular diameter distance, $z$ is the redshift, and $\int n_\textrm{e} n_\textrm{p}$ is the integrated cluster emissivity over the chosen volume of gas in the cluster. The electron density $n_\textrm{e}$ and ion density $n_\textrm{p}$ are related as $n_\textrm{e} = 1.18 n_\textrm{p}$. The electron pressure and entropy are then determined as $P_e = n_\textrm{e} kT$ and $K = kT n_\textrm{e}^{-2/3}$, respectively.

\begin{figure}
    \centering
    \includegraphics[width=\columnwidth]{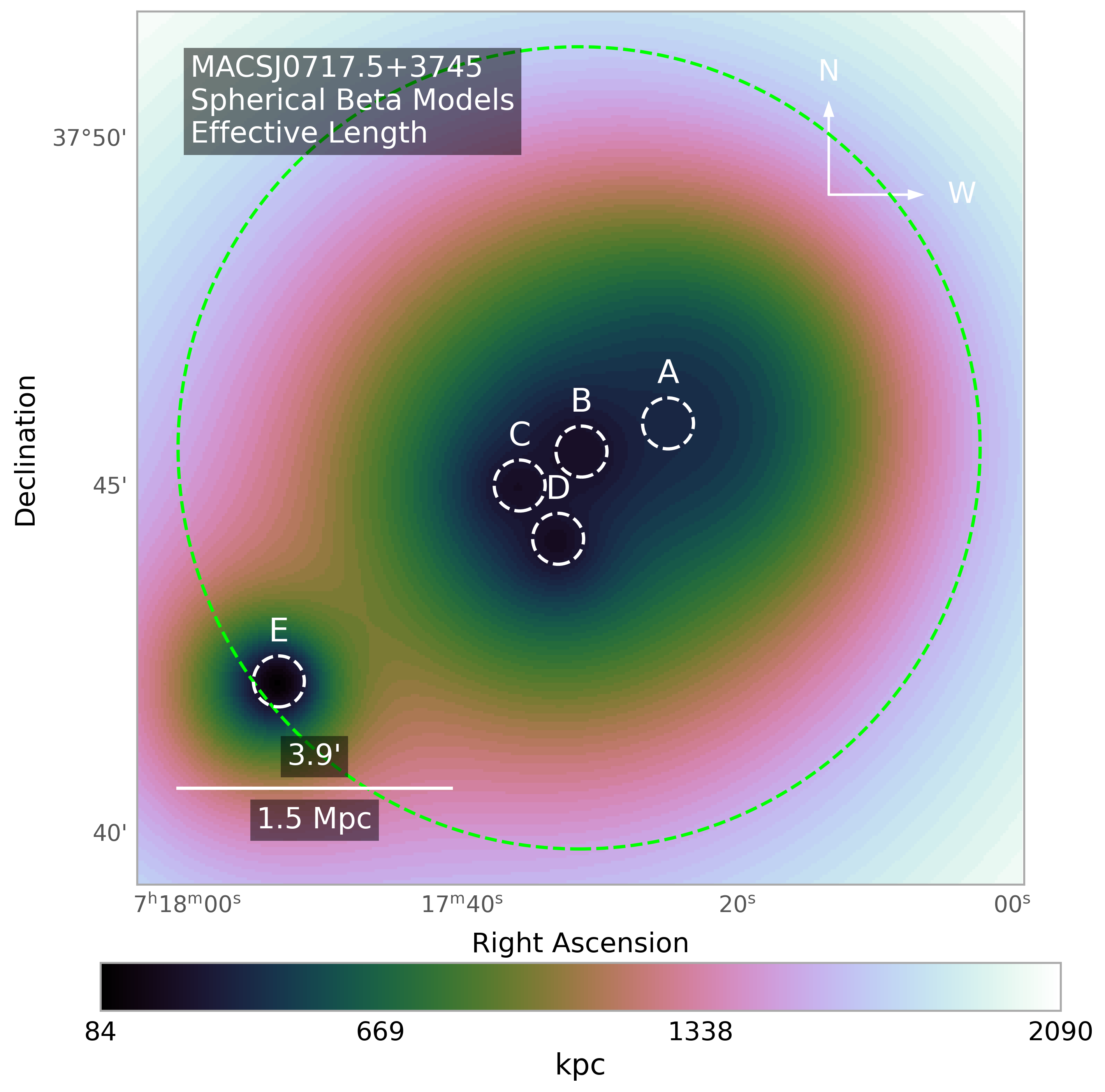}
    \caption{Map of the line of sight effective length derived from the 3D numerical simulation. The five subcluster centers are shown as white circles, and correspond to the different beta models shown in Table~\ref{tab:beta_models}.}
    \label{fig:geometry}
\end{figure}

Given the inherent unknowns and uncertainties associated with the true three-dimensional geometry of \cluster{}, we explore the systematic effects resulting from the simplifying assumption of a constant 1 Mpc line of sight depth in our original thermodynamic maps. To quantify potential biases introduced by this approximation, we implement a geometrical correction derived from a model of the cluster in a 3D box. Specifically, we generated an effective line-of-sight length map using 5 different 3D spherical beta models parameterized according to the lensing-detected subclusters listed in Table~\ref{tab:icm_modeling}, with an additional beta model added to include the group at the end of the filament. The best fit results are given in Table~\ref{tab:beta_models}. The emissivity distribution was then integrated from the center of the 3D simulated cube until 50\% of the total flux along the line-of-sight was reached, shown in Fig.~\ref{fig:geometry}.

\setlength{\tabcolsep}{6.4pt}
\renewcommand{\arraystretch}{1.4}
\begin{table}
\centering
\caption{Best‐fit parameters for five spherical beta‐models determined from Chandra data. The central positions for subclusters A-D are the same shown in Table~\ref{tab:icm_modeling}, and are determined form the strong lensing centers in \citet{Limousin16}, while the center of subcluster E is fixed to the center of the group region.}
\label{tab:beta_models}
\begin{tabular}{llllll}
\toprule
Sub. & Amp. & $r_c$ [kpc] & $\beta$ & RA & Dec \\
\midrule
A & 0.577 & 785 & 1.855 & 07:17:25.0 & +37:45:54.6 \\
B & 1.483 & 279 & 0.896 & 07:17:31.3 & +37:45:30.3 \\
C & 0.851 & 166 & 0.624 & 07:17:35.8 & +37:45:01.0 \\
D & 0.977 & 145 & 0.577 & 07:17:33.0 & +37:44:15.0 \\
E & 0.549 & 69 & 0.495 & 07:17:53.3 & +37:42:11.8 \\
\midrule
\bottomrule
\end{tabular}
\end{table}

Furthermore, we quantify the effect of the geometric correction used in this paper with additional maps of the fractional change and associated sigma change, using the propagated uncertainties derived from spectral fitting for each region, which can be seen in Appendix~\ref{sec:los_estimate} Fig.~\ref{fig:frac_diff} and Fig.~\ref{fig:sigma_diff}.

\begin{figure*}
  \includegraphics[width=\textwidth]{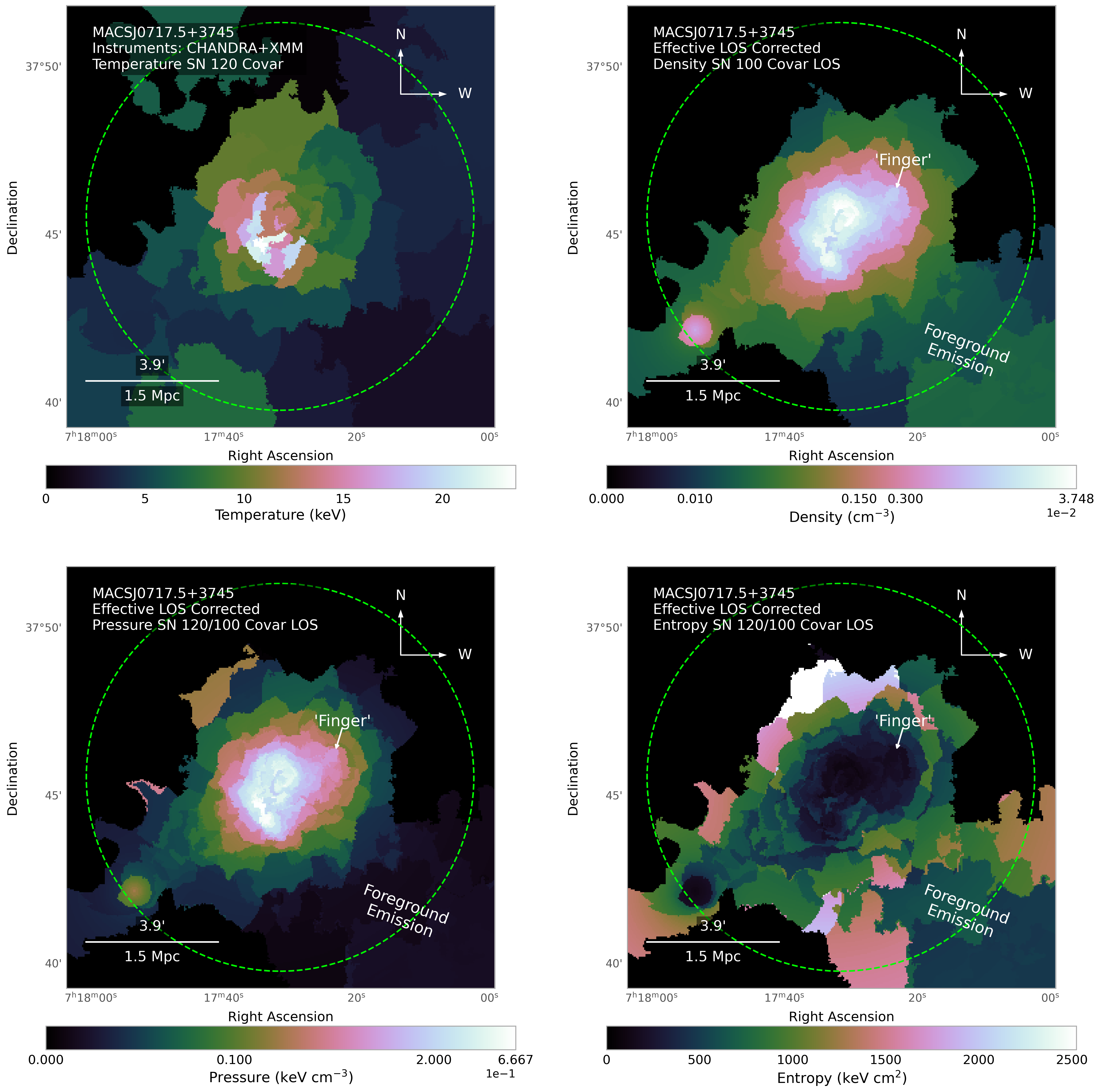}
  \caption{Maps of temperature (\textit{top left}), density (\textit{top right}), pressure (\textit{bottom left}), and entropy (\textit{bottom right}) with corresponding units of keV, cm$^{-3}$, keVcm$^{-3}$, and keVcm$^2$. The temperature and electron density maps were created using binning with a S/N of 120 (14400 counts) and 100 (10000 counts), respectively, while the pressure and entropy maps were produced using the results from the top two S/N maps. Density maps were derived from the normalization fits assuming an effective length determined from the best fit results of 5 beta models, shown in Table~\ref{tab:beta_models}. Spectral fits were made using data from both \chandra{} and \xmm{} for the temperature, and only \chandra{} for the density, each using the CXB best fit results from \rosat{}, \chandra{}, and \xmm{}. Fits for these maps were performed using the most appropriate background model for each region according to the AIC. }\label{fig:thermo_maps}
\end{figure*}

\subsection{Azimuthal Trend-divided Maps}
\begin{figure*}
        \includegraphics[width=\textwidth]{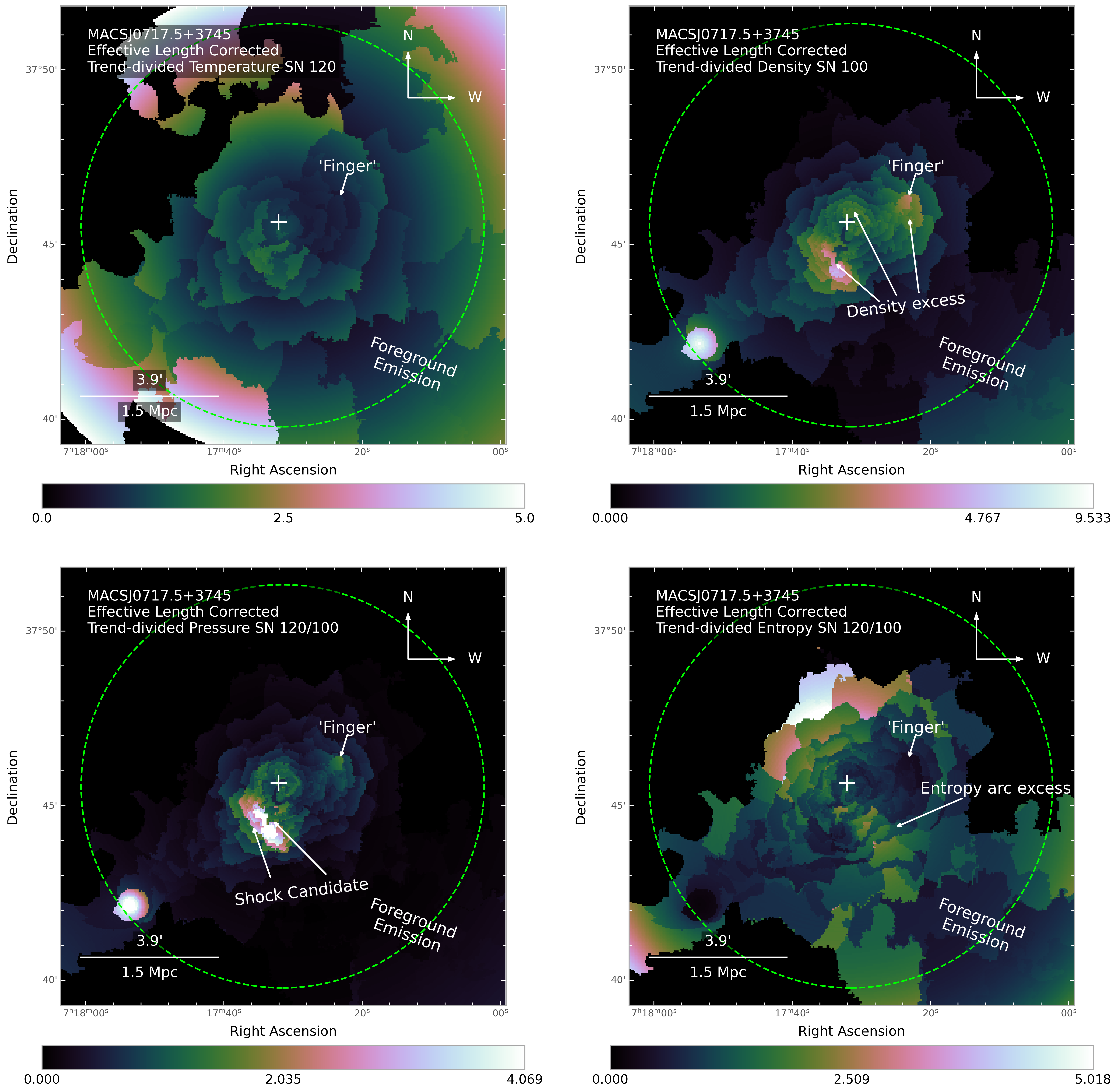}
        \caption{Geometry-corrected, trend-divided maps of temperature (\textit{top left}), density (\textit{top right}), pressure (\textit{bottom left}), and entropy (\textit{bottom right}).\label{fig:trendmaps}}
\end{figure*}

We additionally create trend-divided maps to enhance the small azimuthal variations present in the cluster seen in Fig.~\ref{fig:trendmaps}. We begin by calculating a centroid that is the average between the center of the central contour bin (Bin 0), the position of the peak X-ray surface brightness pixel, and the best fitting results from a beta model fit to the cluster center. We then construct a scatter plot of the radial distribution for each physical quantity from the cluster centroid, and fit for the average trend using a function of the form,
\begin{equation}
f(r) = A (1+ (r/B)^{2})^{(-3C/2)} (1+ (r/D)^{2})^{(-3E/2)}  \,,
\label{eq:trend_model}
\end{equation}
where $r$ is the radial distance from the centroid, and $A$, $B$, $C$, $D$, $E$ are free parameters, following the method described in \citet{ichinohe15}. Since we want to remove the general thermodynamical trends from the cluster, we do not use the full annuli, but instead restrict the radial distribution of physical quantities to the angles corresponding to the same region used to fit for the cluster outskirts. Trend-divided maps are then created for each physical quantity by calculating a residual for every bin, between the original value and the model-predicted value at the given radius. Since these trend-divided maps are unit-less, they are likewise less sensitive to geometric uncertainties imposed by our chosen cluster geometry.

\subsection{Spectroscopic Disentanglement of the Filament Signal}
The spectroscopic fitting of the filamentary region was performed in several steps. We keep the standard assumption that it is a faint signal contaminated by instrumental backgrounds, as well as both astrophysical foreground and background signals; however, we extend the astrophysical description of the background to also include the local atmospheres of nearby systems. The filament lies between \macscluster{} and a group to the southeast, inside their respective $R_{200}$ emission areas, so to properly constrain the temperature and density of the filament, it is first necessary to disentangle the different contributions to the complex signal.

To constrain the cluster and group-related contributions to the filament signal, large arc regions are extracted, approximately at the same radial distance at which the filament structure is displaced from the cluster and group centers, inside, but close to R$_{200}$. These regions are then independently fit using a similar method described in Sec.~\ref{sec:cluster-thermomaps}, using CXB parameter constraints as determined from the optimal CXB fit, and then applying these posteriors as priors to the subsequent fits. The fit results of the CXB + Cluster and CXB + Group regions are shown in the center of Table~\ref{tab:best_fits}.

As we do not know the exact multi-phase composition of the filamentary gas, we try to statistically disentangle the multi-phase nature of the region using Bayesian X-ray Analysis and model selection using the AIC and BIC criteria described in Appendix Sec.~\ref{sec:autobkg}. Table~\ref{tab:aic_comparison} shows the AIC and BIC values when comparing a combination of different spectral models that could effectively describe the filament region. To fit all of the various models, we used only the CXB and filament spectra to fit the temperature and normalization of the filament data while all other parameters were frozen to their best fits. In these data were 3116 data points, and approximately 3114 degrees of freedom, so the AIC and BIC values are fairly close to each other.

We first compared a simple CXB + Filament (\texttt{apec}) model with a CXB + Filament (\texttt{gadem}) model, which applies a gaussian distribution of emission measure versus temperature to an \texttt{apec} model. Despite being often used to fit multi-phase plasmas, \texttt{gadem} was punished by the statistical measure for its extra complexity compared to \texttt{apec}, and was rejected as an option for the more complex model comparisons. We then proceeded to fit a CXB + Cluster + Filament model and a CXB + Group + Filament model, before finally fitting a rather complex model with CXB + Cluster + Group + Filament. This complex model involved the joint-fitting of 49 independent spectra: 13 CXB-associated spectra, and 12 spectra for each of the Cluster, Group, and Filament regions shown in Fig.~\ref{fig:squareboth}.

Based on the different criteria, all of the various \texttt{apec} models were close enough in their likelihoods with each other that they were not outright rejected and theoretically could each be effectively used to describe the data. That being said, the most complicated of the models, which contained the CXB + Cluster + Group + Filament, ended up with the best fit results, despite being punished for its extra complexity. Interesting to note is that the `second-best' model according to the AIC is the CXB + Cluster + Filament, and not the expected CXB + Group + Filament model. 

\begin{figure}
  \centering
  \includegraphics[width=\columnwidth]{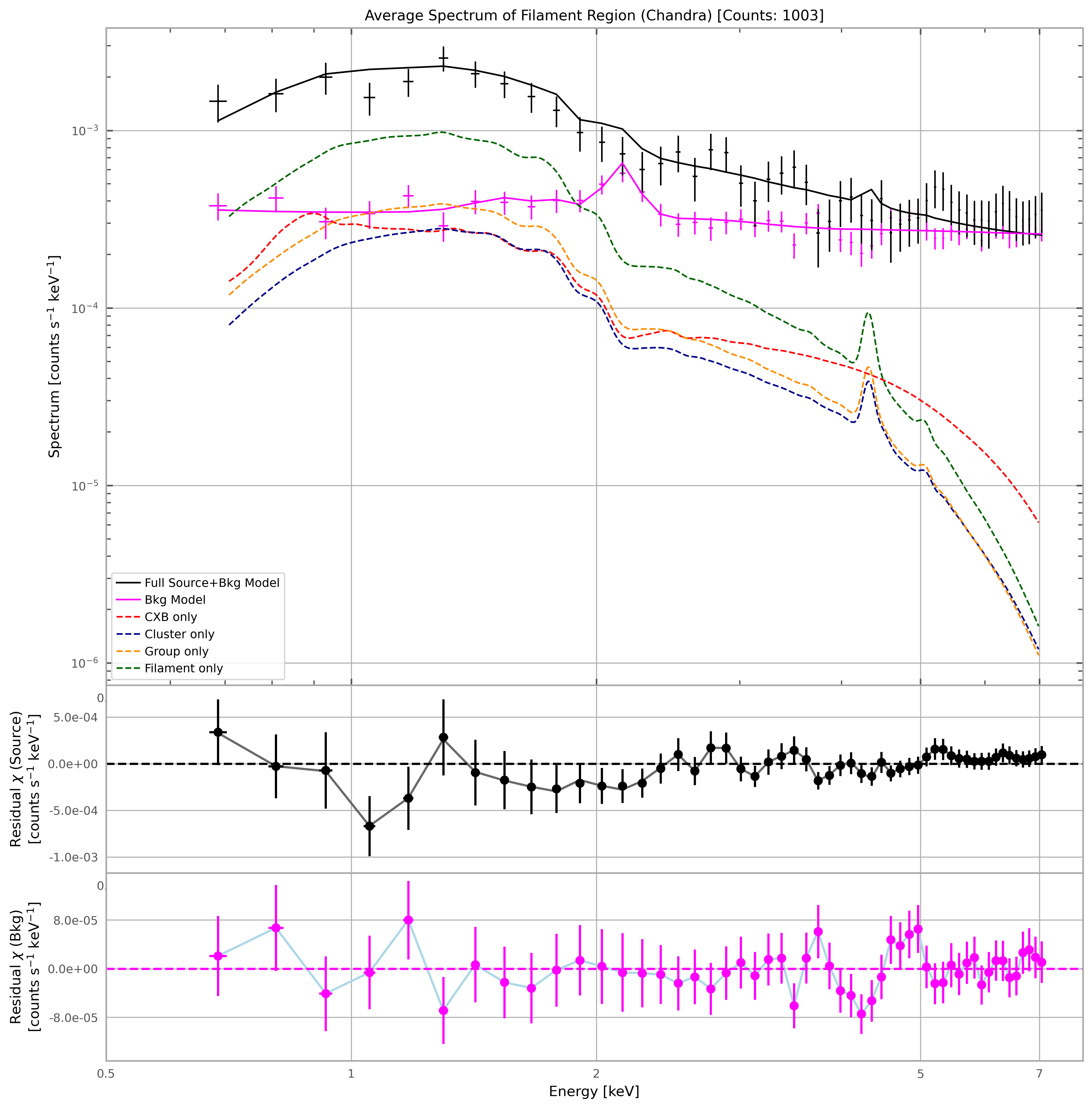}
  \caption{Chandra filament spectra, including all astrophysical and instrumental model components. Spectral data was averaged for each energy bin in all relevant observations, and then re-binned for better visibility. The middle and lower plots respectively correspond to the residuals for the astrophysical source data with respect to the full model, and the instrumental background data and related instrumental background model.}
  \label{fig:filament_spectra_chandra}
\end{figure}

\begin{figure}
  \centering
  \includegraphics[width=\columnwidth]{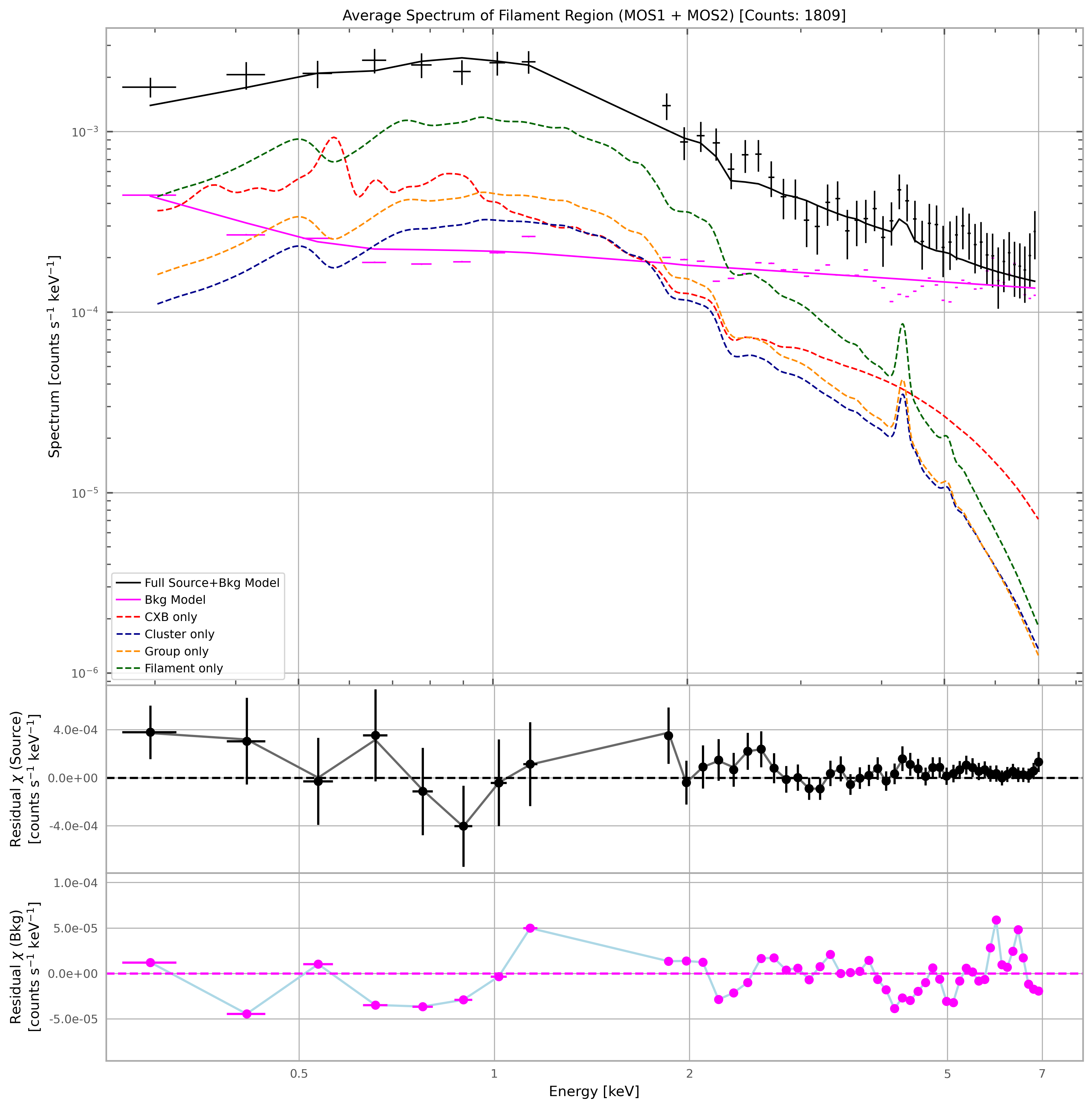}
  \caption{XMM-Newton MOS1+MOS2 filament spectra, including all model components. Spectral data was averaged for each energy bin in all relevant observations, and then re-binned for better visibility. The middle and lower plots respectively correspond to the residuals for the astrophysical source data with respect to the full model, and the instrumental background data and related instrumental background model.}
  \label{fig:filament_spectra_xmm_mos}
\end{figure}

\begin{figure}
  \centering
  \includegraphics[width=\columnwidth]{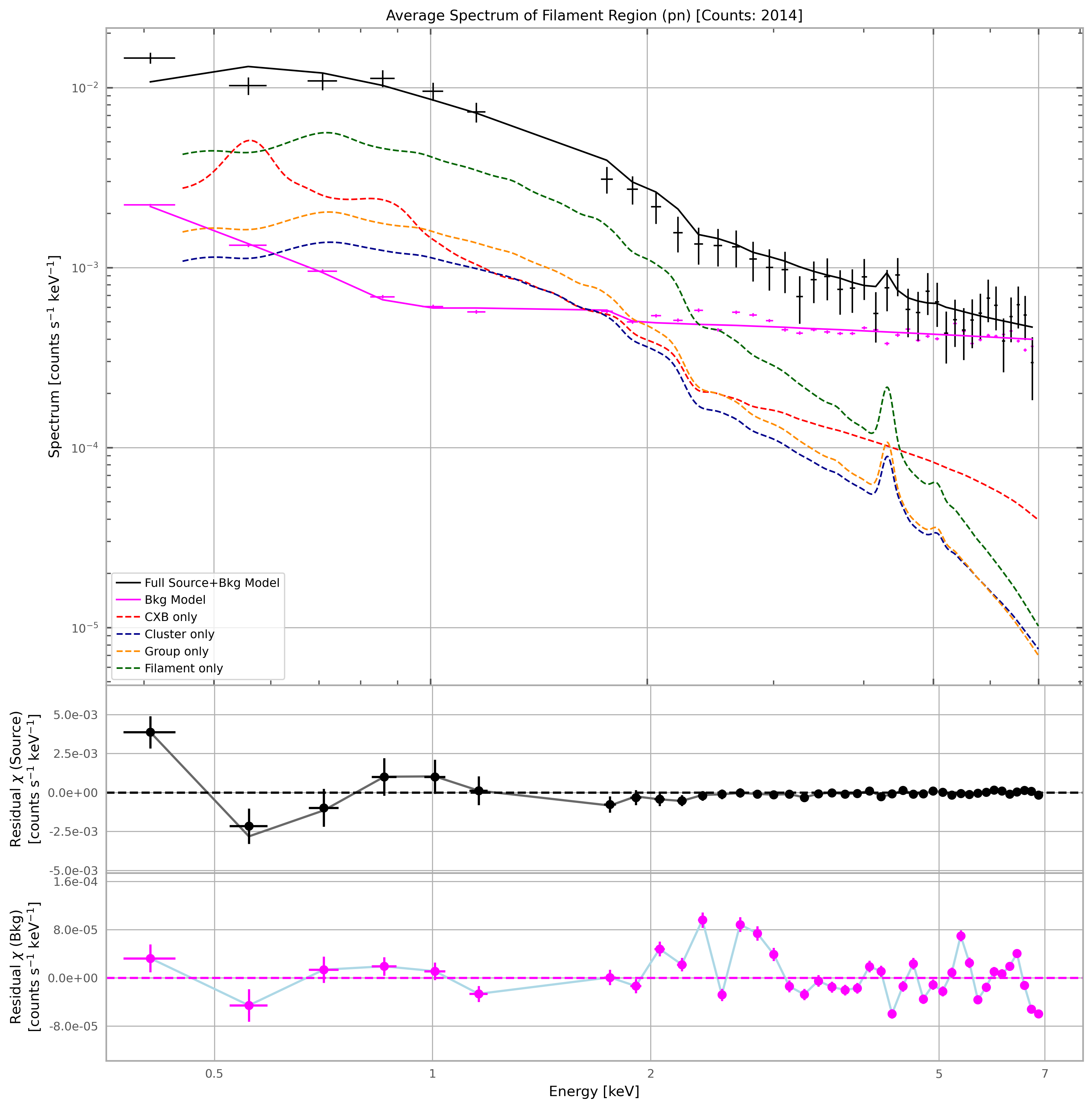}
  \caption{XMM-Newton pn filament spectra, including all model components. Spectral data was averaged for each energy bin in all relevant observations, and then re-binned for better visibility. The middle and lower plots respectively correspond to the residuals for the astrophysical source data with respect to the full model, and the instrumental background data and related instrumental background model.}
  \label{fig:filament_spectra_xmm_pn}
\end{figure}

Figs.~\ref{fig:filament_spectra_chandra},\ref{fig:filament_spectra_xmm_mos},\ref{fig:filament_spectra_xmm_pn} show the source and background spectral data from \chandra{}, \xmm{} MOS1 and MOS2, and \xmm{} pn of the filament region. Besides the full astrophysical and instrumental background models, each plot also shows the individual contributions to the total model from each astrophysical temperature component, along with a residual plot for the spectral data and the full model, and a second residual plot for the instrumental background model. The counts for each instrument are averaged across each energy bin, and then binned for better visualization. As can be seen in the Figures, the contribution from each temperature component was successfully constrained by the fitting. The \xmm{} instrumental background lines between 1.2 and 1.75 keV were not well fit with either a Gaussian, Lorentzian, or Voigt profile, so we ignored that spectral energy range in the spectral fitting. Other features present in the residual are due to the selected background model (in the case of bknpow with no gaussians), or because of cross-calibration between \xmm{} pn, \xmm{} MOS, and \chandra{} data.

\setlength{\tabcolsep}{6.4pt}
\renewcommand{\arraystretch}{1.4}
\begin{table*}
\centering
\caption{The best-fit CXB, Cluster, Group, and Filament parameters from the joint fit using \rosat{}, \chandra{}, and \xmm{} data. The best fit was performed using \textit{cstat} and includes all modelled \chandra{} and \xmm{} instrumental backgrounds. Additional fits of the CXB, Cluster, Group, and Filament parameters excluding the \rosat{} data were performed, with \textit{(cstat)} and without \textit{(wstat)} the instrumental background modelling. \label{tab:best_fits}}
\resizebox{\textwidth}{!}{%
\begin{tabular}{lllrrr}
\toprule
Context & Source & Parameter & No ROSAT (wstat) & No ROSAT (cstat) & Best fit (cstat)\\
\midrule
\multirow{14}{1.2cm}{Cosmic X-ray Background} & \multirow{4}{1.5cm}{Local Hot Bubble} & kT & 0.09 (fixed) & 0.09 (fixed) & $0.108\pm 0.02$\\
    &  & Abundance & 1 (fixed) & 1 (fixed) & 1 (fixed)\\
    &  & Redshift & 0 (fixed) & 0 (fixed) & 0 (fixed)\\
    &  & norm & $(4.5_{-0.2}^{+0.4})\times10^{-6}$ & $(4.5\pm0.1)\times10^{-6}$ & $(7.7\pm0.3)\times10^{-7}$\\
& \multirow{4}{1cm}{Galactic Halo} & kT & $0.22_{-0.01}^{+0.03}$ & $0.209_{-0.004}^{+0.01}$ & $0.156^{+0.002}_{-0.08}$\\
    &  & Abundance & 1 (fixed) & 1 (fixed) & 1 (fixed)\\
    &  & Redshift & 0 (fixed) & 0 (fixed) & 0 (fixed)\\
    &  & norm & $(1.1\pm0.1)\times10^{-6}$ & $(1.1\pm0.04)\times10^{-6}$ & $(2.2^{+3.2}_{-0.05})\times10^{-6}$\\
& \multirow{4}{1cm}{Supervirial Component} & kT & $0.85_{-0.02}^{+0.21}$ & $0.79_{-0.02}^{+0.06}$ & $0.70\pm{-0.01}$\\
    &  & Abundance & 1 (fixed) & 1 (fixed) & 1 (fixed)\\
    &  & Redshift & 0 (fixed) & 0 (fixed) & 0 (fixed)\\
    &  & norm & $(2.5_{-0.4}^{+0.2})\times10^{-7}$ & $(2.8_{-0.3}^{+0.1})\times10^{-7}$ & $(3.6^{+2.6}_{-0.1})\times10^{-7}$\\
& \multirow{2}{1cm}{Unresolved Point Sources} & Photon Index & 1.41 (fixed) & 1.41 (fixed) & 1.41 (fixed)\\
    &  & norm & $(7.6\pm0.1)\times10^{-7}$ & $(7.8\pm0.1)\times10^{-7}$ & $(8.1\pm0.1)\times10^{-07}$\\
\midrule
\multirow{4}{1.2cm}{Cluster} & \multirow{4}{1cm}{Cluster} & kT & ${5.0\pm0.6}$ & ${5.1\pm+0.5}$ & $4.6_{-0.4}^{+0.5}$\\
 &  & Abundance & 0.3 (fixed) & 0.3 (fixed) & 0.3 (fixed) \\
 &  & Redshift & 0.5458 (fixed) & 0.5458 (fixed) & 0.5458 (fixed) \\
 &  & norm & $(7.3\pm0.3)\times10^{-06}$ & $(7.6\pm0.2)\times10^{-06}$ & $(7.83\pm0.02)\times10^{-06}$ \\
\midrule
\multirow{4}{1.2cm}{Group} & \multirow{4}{1cm}{Group} & kT & ${3.6\pm0.6}$ & $4.3\pm0.60$ & $3.8\pm{0.6}$ \\
 &  & Abundance & 0.3 (fixed) & 0.3 (fixed) & 0.3 (fixed) \\
 &  & Redshift & 0.5458 (fixed) & 0.5458 (fixed) & 0.5458 (fixed) \\
 &  & norm & $(1.0\pm0.07)\times10^{-05}$ & $(1.09\pm0.06)\times10^{-05}$ & $(1.12_{-0.06}^{+0.07})\times10^{-05}$ \\
\midrule
\multirow{4}{1.2cm}{Filament} & \multirow{4}{1cm}{Filament} & kT & $2.7_{-0.3}^{+0.4}$ & ${3.2\pm0.4}$ & $3.1_{-0.3}^{+0.6}$ \\
 &  & Abundance & 0.3 (fixed) & 0.3 (fixed) & 0.3 (fixed) \\
 &  & Redshift & 0.5458 (fixed) & 0.5458 (fixed) & 0.5458 (fixed) \\
 &  & norm & $(3.0\pm0.2)\times10^{-05}$ & $(2.9\pm0.2)\times10^{-05}$ & $(3.0\pm0.2)\times10^{-05}$ \\

\bottomrule
\end{tabular}
}
\end{table*}

\setlength{\tabcolsep}{6.4pt}
\renewcommand{\arraystretch}{1.4}
\begin{table*}
\centering
\caption{Comparison of BIC/AIC values for different potential filament models.}
\begin{tabular}{p{6cm}rrc}
\toprule
Model & BIC Value & AIC Value & AIC normalized, log10(Z)\\ 
\midrule
CXB + Filament (gadem) & 4899.90 & 4881.77 & -1.5 \\
CXB + Filament (apec) & 4893.75 & 4881.67 & -0.7 \\
CXB + Group (apec) + Filament (apec) & 4893.71 & 4881.63 & -0.4 \\
CXB + Cluster (apec) + Filament (apec) & 4893.08 & 4880.99 & -0.3\\
CXB + Cluster (apec) + Group (apec) + Filament (apec) & 4892.85 & 4880.76 & 0.0\\
\bottomrule
\end{tabular}
\label{tab:aic_comparison}
\end{table*}

\subsection{The \sz{} Effect}\label{sec:sz_estimates}
Traditionally, the ICM has been studied by using X-ray observations; on the other hand, millimeter-wave astronomy provides an exceptional observational tool to directly probe gas pressure via the \sz{} (SZ) -- \citet{SZ72} -- effect (for a recent review, see \citealt{Tony19}). The \sz{} effect is a spectral distortion in the frequency spectrum of the cosmic microwave background (CMB) radiation due to inverse Compton scattering by a thermal plasma. In practice, it is commonly observed as either a decrement (or increment) in the CMB intensity at frequencies under (or over) 218 GHz along lines of sight through clusters of galaxies, with a linear dependence on the electron density (as opposed to X-ray observations which are dependent on electron density squared). The change in the effective CMB temperature is proportional to the Compton-$y$ parameter, which depends on the Thomson scattering optical depth $\tau_e$, and temperature of the hot electron gas $T_{\rm e}$ as,
\begin{equation}
   y_{tSZ} \equiv \int \frac{k_{\rm B} T_{\rm e}}{m_{\rm e} c^2}d\tau_{\rm e} = \int \frac{k_{\rm B} T_{\rm e}}{m_{\rm e}c^2}n_{\rm e}\sigma_Tdl = \frac{\sigma_T}{m_{\rm e}c^2}\int P_{\rm e} dl
   \label{eq.sz_eq}
\end{equation}
where the Thompson cross section, $P_{\rm e} = n_{\rm e} k_{\rm B} T_{\rm e}$ is the pressure due to the electrons integrated over the line of sight distance \citep{Tony19}. 

There can additionally be another spectral distortion to the CMB spectrum due to the Doppler effect, known as the kinetic \sz{} effect, caused by the cluster bulk velocity along the line of sight on the scattered CMB photons, described by,
\begin{equation}
   y_{kSZ} = \sigma_T \int \frac{-v_z}{c}n_edl \equiv \frac{-v_{\rm z}}{c}\tau_{\rm e}\, . 
\end{equation}
where $v_z$ is the cluster bulk velocity (positive for receding clusters), and $\tau_e$ is the electron optical depth. The observed change of specific intensity $\Delta I_{\nu}$, can be expressed with respect to the CMB intensity $I_0$ as a function of the observed frequency $\nu$, given by
\begin{equation}
    \frac{\Delta I_{\nu}}{I_0} = f(\nu, T_e)y_{tSZ} + g(\nu, T_e, v_z)t_{kSZ}\,,
\end{equation}
where $I_0 = \frac{2(k_BT_{\rm{CMB}})^3}{(hc)^2} = 270.33 \left[\frac{T_{\rm{CMB}}}{2.7255{\rm{K}}}\right] \rm{MJy/sr}$, and $f(x, T_e)$ describes the characteristic tSZ effect frequency dependence, and $g(\nu, T_e, v_z)$ describes the spectral dependence of the kSZ effect in the non-relativistic regime, given by \citet{Birkinshaw99} as,

\begin{equation}
    f(\nu, T_e) = -\frac{x^4 e^x}{(e^x - 1)^2} \left( x \coth \left(\frac{x}{2}\right) - 4 \right) \left( 1 + \delta_{\rm{tSZ}}(x, T_e) \right)\, ,
\end{equation}
\begin{equation}
    g(x, v_z, T_e) = \frac{x^4 e^x}{(e^x - 1)^2} \left( 1 + \delta_{\rm{kSZ}}(x, v_z, T_e) \right)\, ,
\end{equation}
where $x = \frac{h\nu}{k_B T_{\rm{CMB}}} \approx \nu/56.8{\rm{GHz}}$ is the dimensionless frequency, $h$ is the Planck constant, $\nu$ is the observed frequency, $k_B$ is the Boltzmann constant, and $T_{\rm{CMB}} = 2.7255 {\rm{K}}$ is the CMB temperature. The $\delta_{\rm{tSZ}}(x, T_e)$ and $\delta_{\rm{kSZ}}(x, v_z, T_e)$ terms correspond to the relativistic corrections for the tSZ and kSZ effects, and are dependent on the observing frequency and $T_e$, and in the kSZ case, also the line of sight velocity. The relativistic corrections for the tSZ are computed using the work of \citet{ItohNozawa03}, while the kSZ corrections are computed using the analytical formula in \citet{Nozawa06}. Practically, these relativistic corrections are computed from tabulated values using SZpack, whereby the corrections are on the 10\% level for these 18 keV temperatures \citep{szpack,szpack2}.

We introduce here the previously unpublished \mustang{} images in Fig.~\ref{fig:sz} and Fig.~\ref{sec:mustang_fov}. The \mustang{} data was corrected for the kinetic and relativistic SZ effect following a constructed model based on the best fit results of the `F2' model from \citet{Adam2017b}; this model construction is described in detail in Sec.\ref{sec:icm_modelling}.

\subsection{Correction of kinetic \sz{} via ICM Physical Modeling}\label{sec:icm_modelling}

\begin{figure*}
    \centering
    \includegraphics[width=\textwidth]{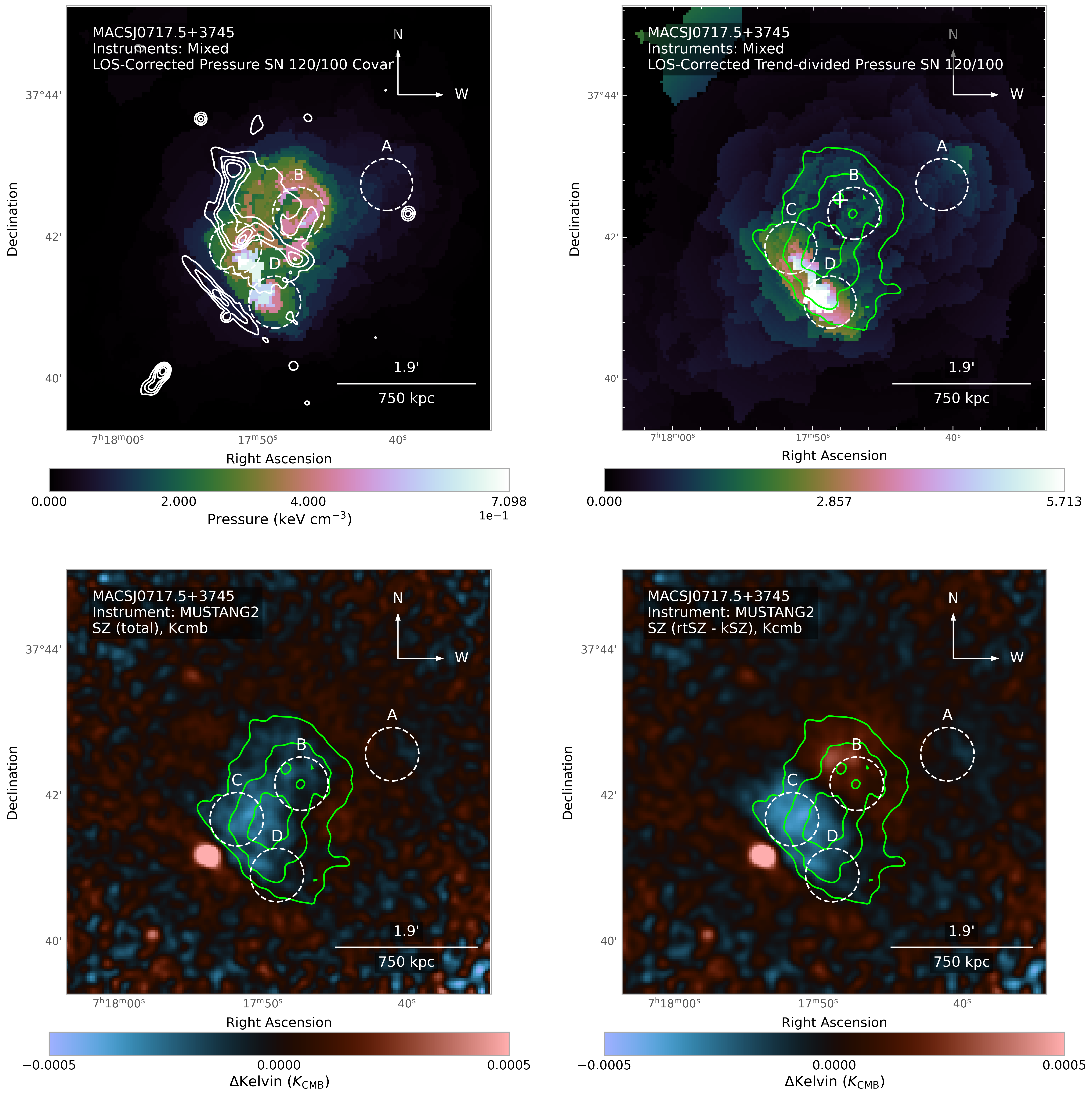}
    \caption{Comparison plots between the geometry-corrected X-ray pressure map (\textit{top left}), the geometry-corrected  trend-divided X-ray pressure map (\textit{top right}), followed by \sz{} \mustang{} data (\textit{bottom left}), next to a kinetic and relativistic SZ corrected version of the data, shown here in Kelvin$_{\rm{CMB}}$ units (\textit{bottom right}). The white contours are from the radio images seen in \citet{VanWeeren2017}, and the green contours corresponds to the \mustang{} data in the bottom left. The circles represent the strong lensing subcluster centers given in \citet{Limousin16} which were used to reproduce the kSZ+rSZ model used in \citet{Adam2017b}, which are displayed in Table~\ref{tab:icm_modeling}. The full FOV of the \mustang{} data can be seen in Appendix Sec.~\ref{sec:mustang_fov}.} \label{fig:sz}
\end{figure*}

\citet{TonySZ_macs} and \citet{Sayers2013} both explored \macscluster{} in the \sz{}, making the first discovery of a cluster system with a kinetic SZ component. \citet{Adam2017a} later made spatially resolved kinetic SZ maps of \cluster{} with the New IRAM KIDs Array (NIKA) on the IRAM 30m telescope, followed by a joint X-ray/SZ study of the projected gas temperature distribution \citep{Adam2017b}. Recently, \citet{Adam2024} also used NIKA data to probe the turbulent gas motions in \cluster{}, finding a kinetic to kinetic plus thermal pressure fraction $P_{\rm {kin}}/P_{\rm {kin+th}}$ of around 20\% and an energy injection scale of around 800 kpc. 

Because \mustang{} is dominated by the tSZ and limited to a single frequency $\sim90\rm{GHz}$, it does not have much constraining power for the kSZ signal. As we can not constrain our own kSZ model, we instead opt to build a kSZ model for \mustang{}. To make the correction, we use the same model parameters as `F2' constrained in \citet{Adam2017b}, which used the NIKA data and extra information from X-ray imaging to break degeneracies between $v_z$ and $\tau_e$. We start with the fact that previous strong lensing observations in \citet{Limousin16} have shown that there are at least four subclusters present in \cluster{}, listed in Table~\ref{tab:icm_modeling}, and shown as circles in Fig.~\ref{fig:sz}. Due to the linear dependence of $n_e$ in the SZ data, the 3D gas distribution is built as a sum of each independent subcluster, where each subcluster is modeled with a spherically symmetric $\beta$-model,

\begin{equation}
    n_e(r) = n_{e0}\left[1 + \left(\frac{r}{r_c}\right)^2\right]^{-3\beta /2}\, ,
\end{equation}
and $n_{e0}$ is the central electron density, $r_c$ is the core radius, and $\beta$ is the slope of the radial profile \citep{beta_model}. Each subcluster additionally is assumed to have isothermal gas temperatures, determined using 30 arcsecond regions from \xmm{} in \citet{Adam2017b}, and constant bulk velocities, all given in Table~\ref{tab:icm_modeling}. This gas distribution is then analytically integrated along the line of sight for each subcluster as,

\begin{equation}
    \tau_e \equiv \sigma_T\int_{-\infty}^{+\infty} n_e dl = \sqrt{\pi} \, \sigma_T \, n_{e0} r_c
\frac{\Gamma\left(\frac{3}{2} \beta - \frac{1}{2} \right)}{\Gamma\left(\frac{3}{2} \beta \right)}
\left[ 1 + \left( \frac{R}{r_c} \right)^2 \right]^{\frac{1}{2} - \frac{3}{2} \beta}\, ,
\end{equation}
where $\Gamma$ represents the Gamma function, and $R$ is the projected radius from the cluster center. This can then be used to create the tSZ and kSZ models used to compare with the \mustang{} surface brightness maps, given by,

\begin{align}
    \frac{\Delta I_{\nu}^{\rm{model}}}{I_0} &= \sigma_T \sum_i f_{\nu}(T_x^{(i)}) \frac{k_B T_x^{(i)}}{m_e c^2} \int n_e^{(i)} dl \notag \\
&+ \sigma_T \sum_i g_{\nu}(T_x^{(i)}) \frac{v_z^{(i)}}{c} \int n_e^{(i)} dl\, ,
\end{align}
where $T_x$ and $v_z$ are scalar quantities for each subcluster.

Fig.~\ref{fig:sz} shows a comparison plot between a zoomed in view of the geometry-corrected X-ray pressure and the geometry-corrected, trend-divided pressure maps (\textit{top}), followed by \sz{} \mustang{} data, and a kSZ and rSZ corrected version of the data. The white box corresponds to the FOV of the \mustang{} data, and is the same as the thermodynamical maps shown in Fig.~\ref{fig:thermo_maps} and Fig.~\ref{fig:trendmaps}, while the new smaller green box shows the zoomed-in region in the two top plots. The white contours are the radio images in \citet{VanWeeren2017}, and the circles represent the subcluster centers given in Table~\ref{tab:beta_models} which were used for the geometric correction of the maps and Table~\ref{tab:icm_modeling} used in the kSZ model.

There is a relatively close agreement between the X-ray pressure peak and the SZ peak, with an offset corresponding to around 0.25 arcminutes, or around 100 kpc at this redshift, but this estimate is biased due to the resolution of the \mustang{} data and the size and shape of the corresponding bin from the X-ray pressure map. As can be seen from the trend-divided pressure map, there is also a clear offset between the radio relic emission and the X-ray pressure peak, likely due to the nature of the integrated X-ray emission and projection effects mentioned in \citet{VanWeeren2017}. Despite the large uncertainties and differences in the assumptions associated with the kSZ model, our results are consistent with the previously published results in \citet{Adam2017b}. Considering the noise in the \mustang{} data, shown in Appendix Sec.~\ref{sec:mustang_fov}, the kSZ correction applies the most significant correction around subcluster B of around 2.9$\sigma$. The correction is less significant at the other subclusters, specifically -0.31$\sigma$, -0.96$\sigma$, and -0.28$\sigma$ for subclusters A, C, and D, respectively.

As mentioned previously, the present work does not explore pressure fluctuations \citep[see][]{tsz2016, tsz2023, tsz2024}, but we note that deeper SZ observations with, for example, \mustang{}, could improve upon the constraints for \cluster{} presented in \citet{Adam2024}.

\setlength{\tabcolsep}{6.4pt}
\renewcommand{\arraystretch}{1.4}
\begin{table*}
\centering
\caption{Parameters used for creation of the kSZ and rSZ corrected map following the best fit model `F2' given in \citet{Adam2017b}, the four subcluster centers are given by the strong lensing centers in \citet{Limousin16}. These subcluster centers are shown in Fig.~\ref{fig:sz} as 22 arcsecond radius regions for visualization purposes.}
\begin{tabular}{cccccccc}
\toprule
Subcluster & $n_{e0}$ [cm$^{-3}$] & $r_c$ [kpc] & beta & $T_e$ [keV] & $v_z$ [km/s] & R.A. & Dec.\\ 
\midrule
A & 0.003452 & 408.953 & 2.26335 & 7.56708 & 0 (not fit) & +07:17:25.0 & +37:45:54.6\\
B & 0.0105003 & 472.866 & 2.46193 & 10.6967 & 2028.92 & +07:17:31.3 & +37:45:30.3\\
C & 0.0124379 & 289.809 & 2.39821 & 16.3905 & -4820.22 & +07:17:35.8 & +37:45:01.0\\
D & 0.00948142 & 466.939 & 2.42118 & 12.5475 & 882.430 & +07:17:33.0 & +37:44:15.0\\
\bottomrule
\end{tabular}
\label{tab:icm_modeling}
\end{table*}

\section{Discussion}\label{sec:discussion}

\subsection{Thermodynamical Maps}\label{sec:thermo_discussion}
The temperature map on the top left of Fig.~\ref{fig:thermo_maps}, shows an exceptionally hot cluster center, at around 24$\pm 4$ 
keV, with a temperature dropping to 9.8$\pm 0.8$ keV in the south and to 13$\pm 3$ keV in the north-northwest. North of the temperature peak, and slightly west of the radio contours (see Fig.~\ref{fig:sz}), we find a $\sim$ 20 keV elongated sub-structure; likewise, to the south and west of the peak, we see another $\sim$20 keV sub-structure. 

Farther out towards the outskirts, in the north-northeast direction, we see a spatially large region with a temperature of $\sim$ 10 keV. To the west of \macscluster{}, after we pass the $\sim$ 10 keV region, we observe $\sim5$ keV plasma surrounded by 6-7 keV plasma. Results approaching the $R_{200}$ are probably biased by foreground structures and extended emissions to the west. In the south-southeast direction, we see a filamentary structure connecting the cluster with a group of galaxies. The filament temperature is reported in Table~\ref{tab:best_fits}. In the map, the group itself has a temperature of 3.61$\pm 0.17$ keV, whereas the temperature reported in Table~\ref{tab:best_fits} gives the group $R_{200}$ temperature. Based on previous lensing studies, it was determined that the filament continues farther out in this direction, even branching off into two separate filaments \citep{Jauzac2018}. Unfortunately, our X-ray data do not allow us to detect the X-ray emission associated with this part of the filamentary structure.

The measured density map traces the X-ray surface brightness image. We can also clearly see the density excess of the filament connecting the cluster and the group. Additionally, the density map appears to trace the surface brightness excess at the `finger' structure to the north-northwest described by \citet{VanWeeren2017}. 

The large temperature and pressure discontinuities that we observe in the center of the cluster indicate a potential presence of a shock candidate.

In particular, the pressure discontinuity to the southeast of the haloes C and D, (see Fig.~\ref{fig:sz}) indicates that we may be looking at a shock candidate. This would also be consistent with the radio emission reported by \citet{VanWeeren2017}. Previous discussions about the nature of these radio structures in \citet{VanWeeren2017} proposed the hypothesis that these are radio relics created as a result of subcluster merger activity, which are seen at some peculiar projected geometry. Based on our temperature map, we estimate the Mach number for the shock candidate somewhere between $\mathcal{M} = 1.7 \pm 0.3$ and $\mathcal{M} = 2.0 \pm 0.3$. \citet{VanWeeren2017} gives an estimate of the Mach number of the shock derived from the radio spectral index of $\mathcal{M} = 2.7$, which is not significantly different from our X-ray values, considering the various uncertainties. Alternatively, the observed pressure peak could be due to the gravitational potential peak of the two haloes (C and D). However, this scenario appears to be in tension with the observed steep temperature and pressure gradients.

For a cluster in hydrostatic equilibrium, we expect to see an azimuthally symmetric and radially monotonously increasing entropy. Any departure from such a distribution is either due to the infall of subgroups hosting lower entropy material, uplift from the cluster center due to AGN feedback, or shock heating. The entropy map shows that the group in the southeast has a relatively low entropy near the center, which gradually increases with radius. The entropy of the X-ray bright bridge/filament appears lower than that of the surrounding ambient ICM at the same radius, which would be consistent with the potential infall of stripped material. At the position of the linear radio source, we see indications of an entropy discontinuity consistent with the presence of a shock front. This linear bar region to the south of the cluster center is also cospatial with the temperature/pressure peak and the associated X-ray shock features. Meanwhile, the characteristic 'V' shape, which is present in the cluster center, seems to be associated with the colder region just north of the shock. The complex nature of the past subcluster merger activity and unknown geometry gives these shock discontinuities asymmetric and complex physical structures in projection. 

Curious features also become apparent in the various trend-divided thermodynamical maps. In the trend-divided density map, besides the filamentary region to the south-west, three subclusters near the cluster center show excess density, specifically the bar region associated with the shock front, the 'V' shaped emission region at the cluster center, and the so-called 'finger' region to the north-west of the cluster. These regions also appear overpressured and are associated with entropy decrements in the trend-divided pressure and entropy maps. New interesting features, now visible in the trend-divided entropy map, are two high entropy regions originating from the cluster center and going to the northeast and southwest. Looking at the radio contours present in Fig.~\ref{fig:sz}, the trend of this high entropy feature near the cluster center seems to trace the radio shock reported by \citet{VanWeeren2017}, however, near the outskirts of the cluster, it may be associated with the foreground structures present in the west of the cluster system. 

\subsection{The Filament}
The area of our selected rectangular filament region is 1.2 arcmin$^2$, assuming a projected geometry, we calculate a volume of an elongated rectangular cuboid inclined near the line of sight based on the geometry proposed by \citet{Jauzac2012}, where the assumed projected line of sight depth was determined to be $\sim 4000\,{\rm kpc}$.

The filament temperature of $3.1^{+0.6}_{-0.3}$ keV is lower than the cluster and the group temperatures at the same radius. The inferred filament density based on the assumed volume is $(3.78\pm0.05)\times10^{-4}\,{\rm cm^{-3}}$. The filament has a lower entropy than the cluster and group outskirts. This can also be seen in the entropy map in the bottom right of Fig.~\ref{fig:thermo_maps}, where a low entropy bridge is surrounded by a gas of higher entropy.

At the cluster redshift, the critical density of the Universe is $1.7\times10^{-29}$ g cm$^{-3}$. If we assume that the total baryon density is 0.044 of the critical density of the Universe and that the baryon mass fraction is 0.15 of the total mass density \citep{kirkman2003}, we get an overdensity of our filament of 729 relative to the mean matter density of the Universe, and an overdensity of 401 relative to the critical density. Using the same volume, the baryonic mass of the filament is $\sim6.1\times10^{12}~\rm M_\odot$. Assuming that the baryon fraction of the filament equals the mean cosmic value, its total mass would be $\sim7.6\times10^{13}~\rm M_{\odot}$.

The inferred temperature of the filament region is relatively high compared to what would be expected of a WHIM filament. The ambiguity from the projection of the filament structure provides two potential hypotheses for this. Firstly, the increased temperature might be the result of compression as the cluster and group atmospheres interact with each other. This hypothesis is supported by the fact that the calculated density for the filament region is also relatively high for what is expected from a typical WHIM filament; so if it is a true WHIM filament, it should be compressed by the merging atmospheres. Alternatively, the excess temperature could be also explained by shock heating from the WHIM accretion in the outskirts of the group atmosphere, similar to the proposed {\normalfont \itshape external accretion shocks} \citep{Ha2018, Gu19}. Lensing measurements of the filament structure, which reveal that it extends further to the southeast of the group, support this scenario. These types of shocks should, in theory, have high Mach numbers ($\mathcal{M} \sim 10-100$), but have not yet been confirmed observationally since they are present in regions with very low X-ray surface brightness. Alternatively, the cooler component can be due to stripped low entropy gas from groups and galaxies seeding gas inhomogeneities in the direction of the large scale structure filament. A similar scenario was proposed for Abell 85 by \citet{ichinohe15}. 

\begin{figure}[t!]
\centering
\includegraphics[width=\columnwidth]{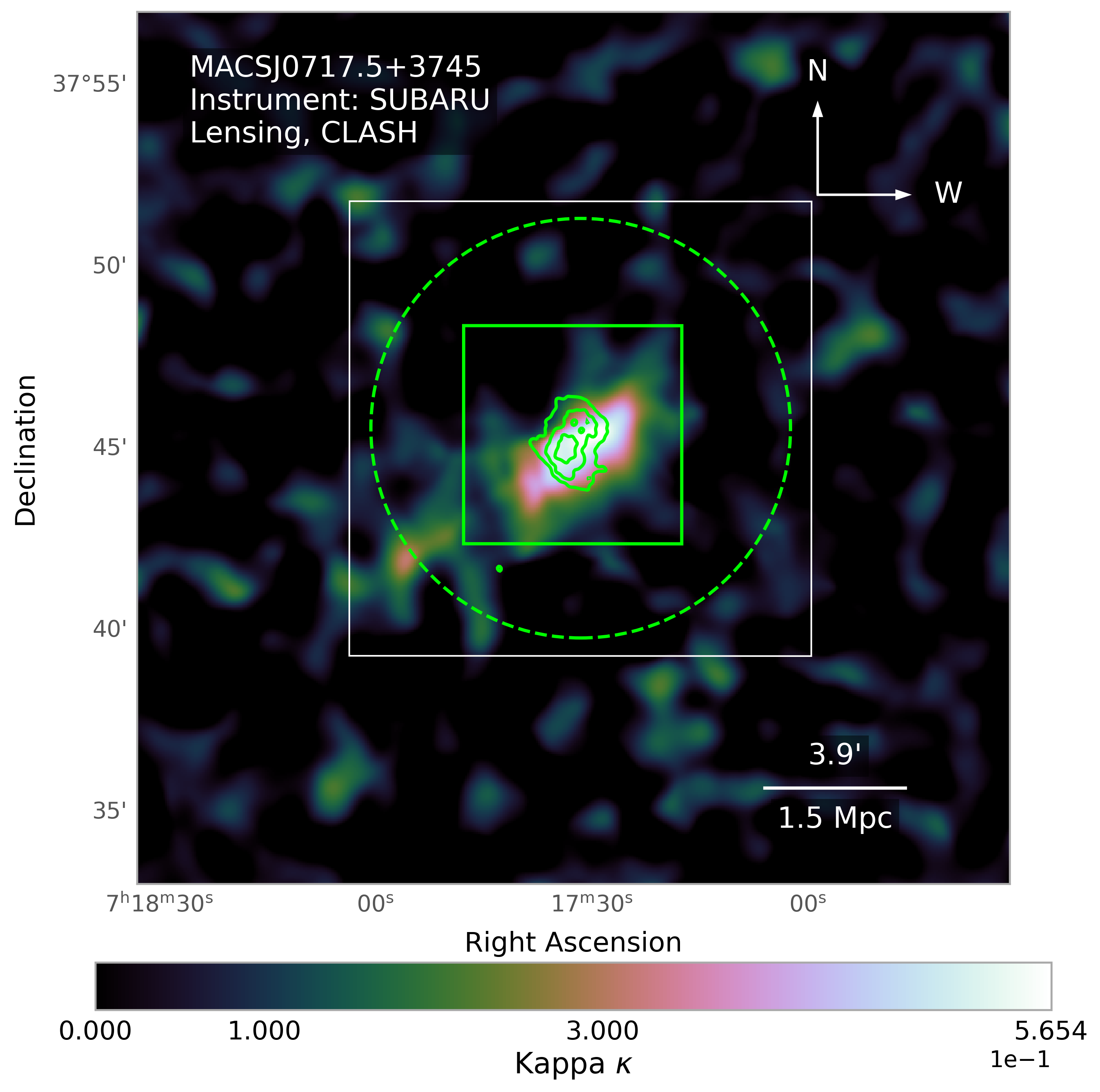}
\caption{Weak lensing map from the CLASH survey \citep{clash}. The white box corresponds to the same FOV as the \mustang{} data and the other thermodynamical maps. The green box corresponds to the zoom-in region shown in Fig.~\ref{fig:sz}.}
\label{fig:lensing}
\end{figure}

Using the wide-field Subaru weak-lensing observations of \citet{clash} to measure the mean surface mass density $\overline{\Sigma}$ within the filamentary region, we can provide an upper limit to the total mass of the filamentary structure. We note that the mass map shown in Fig.~\ref{fig:lensing} was derived using Gaussian smoothing for visualization purposes \citep{clash}, which is not suitable for accurate mass estimation. To this end, we employ the pixelized surface mass density map $\Sigma(\vec\theta_i)$ in the cluster field and its covariance matrix $C_{ij}$ from \citet{umetsu2018}, derived from the combination of two-dimensional reduced shear and azimuthally averaged magnification-bias constraints using the cluster lensing mass inversion 2D ({\sc clumi}-2D) code. Combining complementary shear and magnification measurements, this method reconstructs the underlying surface mass density field around the lens, effectively breaking the mass-sheet degeneracy \citep[see][for a review of cluster weak lensing]{umetsu2020}. The mass map $\Sigma(\vec\theta_i)$ covers a field of $24\times 24$~arcmin$^2$ with $N_\mathrm{pix}=48\times 48$ pixels centered on the cluster. We find $\overline{\Sigma}=(3.7\pm 1.4)\times 10^{14}~\rm M_\odot$~Mpc$^{-2}$ using an optimal (minimum variance) estimator, $\overline{\Sigma}=[A^t C^{-1}A]^{-1}A^t C^{-1}\Sigma$ \citep[Appendix C of][]{umetsu2015}, where $A$ is an $N_\mathrm{pix}\times 1$ mapping matrix whose elements $A_i$ are unity for those pixels lying inside the filamentary region and zero otherwise. The uncertainty for $\overline{\Sigma}$ is given by $\sigma(\overline{\Sigma})=(A^tC^{-1}A)^{-1/2}$. This translates into a projected mass inside the filament region of $(6.8 \pm 2.7) \times10^{13} \, \rm M_{\odot}$. These large uncertainties are reasonable, given that the weak lensing-based mass reconstruction has a pixel scale of 0.5~arc\-min ($\sim 190$~kpc), which corresponds to only 7 pixels inside the filament region.

Taking these gas mass estimates ($4.2\times10^{12}~\rm M_\odot$) and the projected total mass measurements ($6.8\pm 2.7\times10^{13}~\rm M_\odot$) at face-value, the gas mass fraction of the filament is 4--10\%. This range is comparable to the upper limit of 9\% determined for the filament in Abell 222/223 \citep{dietrich2012} and within the range of 5--10\% inferred for the filament system around Abell 2744 \citep{eckert2015}.  

\section{Summary}
\macscluster{} is possibly one of the most dynamically active clusters in our Universe, making it also one of the most challenging galaxy clusters to study in the X-ray, owing to its geometric complexity, extremely high central temperature, as well as the many projected merging systems. Its rich radio morphology, X-ray substructures, and nearby filamentary structures have also paved the way for many interesting papers across different wavelengths.

In this paper, we presented results from deep X-ray observations of the merging cluster \macscluster{} using a novel approach to dynamically model the instrumental and astrophysical backgrounds, as well as a new method to jointly fit \chandra{}, \xmm{}, and \rosat{} data from the entire FOV. We apply this method to constrain the CXB and instrumental background components, before fitting the respective regions, in an attempt to push the imaging and spectroscopic capabilities of these observatories to their limit. We provide new thermodynamical maps as far out to $R_{200}$ as possible before  using joint modeling of all available data, providing an order of magnitude improvement in spatial resolution with respect to the previous spectroscopic maps. We also present a novel use of statistical model comparison methods to disentangle the complex spectral emission from several overlapping X-ray components using Bayesian analysis tools, finding a new temperature component in the filamentary structure to the S-SE of the cluster. We also present a sensitive  map of the Sunyaev-Zeldovich Effect decrement $10''$ resolution from MUSTANG-2 on the GBT.

To summarise the results:
\begin{itemize}
    \item Thermodynamic maps produced from \chandra{}, \xmm{}, and \rosat{} data of \macscluster{}, using a new method for modelling both the astrophysical and instrumental backgrounds reveal a complicated ICM structure with several subsystems. The region south-southeast of the cluster center shows the presence of exceptionally hot 24$\pm 4$ keV gas and pressure discontinuities in the cluster core indicate the presence of shocks. To the south-southeast, we detect the X-ray emission of a large-scale structure filament connecting the cluster with a group.

    \item We describe \macscluster{} by creating a 3D cluster model assuming five spherical beta models. The geometric correction is based on the four strong lensing centers previously reported in the literature and the group at the end of the filament structure. Using this model, we calculate the effective line-of-sight length, which is used to calculate the volume to determine the density, pressure, and entropy in the thermodynamic maps.

    \item Trend-divided maps reveal regions of excess density that are related to overpressured regions with complex entropy structure. These regions appear approximately co-aligned with the radio relic emission reported by \citet{VanWeeren2017}. The filament structure and other subclusters are more clearly visible after removing the cluster average trend for each physical quantity.

    \item The temperature peak of 24$\pm 4$ keV is also the pressure peak of the cluster and is spatially offset from the \sz{} peaks by 0.25 arcmin, around 100 kpc, located approximately halfway between the two lensing mass peaks. We report a range for the Mach numbers for the potential shock candidate between $\mathcal{M} = 1.7 \pm 0.3$ and $\mathcal{M} = 2.0 \pm 0.3$.

    \item We present the first \mustang{} SZ maps of \macscluster{}, which were corrected for the kinetic and relativistic SZ effects using a model based on the best fit results from \citet{Adam2017b}. The SZ maps show a close agreement with the X-ray pressure peak, and the kSZ model corrections are consistent with previously published work.

    \item Bayesian X-ray Analysis methods were used to disentangle different projected spectral signatures for the filament structure, with Akaike and Bayes criteria being used to select the most appropriate model to describe the various temperature components. The statistical measures indicate that the most complex model, involving CXB+Cluster+Group+Filament components is the best model to represent the filament structure. We report X-ray filament temperature of $3.1_{-0.3}^{+0.6}$ keV and density $3.8\pm0.1\times10^{-4}\,{\rm cm^{-3}}$, corresponding to an overdensity of the filament of 729 relative to the mean matter density of the Universe, and an overdensity of 401 relative to the critical density. We estimate the baryonic mass of the filament to be $\sim6.1\times10^{12}~\rm M_\odot$, while its total projected weak lensing measured mass is $\sim6.8\pm2.7\times10^{13}~\rm M_\odot$, indicating a hot baryon fraction of 4–10\%.
  \end{itemize}

  \section*{Acknowledgements}
  Firstly, JPB would like to thank the reviewer for his/her comments to the different drafts of the report. The implemented changes to the paper have added much to the scientific value and potential impact of the results. JPB would also like to give a big thank you to the members of the Chandra \textsc{Ciao/Sherpa} help desk and the various calibration and instrument scientists for all of their help and advice. Special thanks is to be given to Nick Lee, for spending a year and over 120 emails helping me troubleshoot the development of the instrumental background fitting procedure, and for helping understand the many discovered bugs regarding the AREASCAL keywords and the background scaling in \textsc{Sherpa}. Special thanks is also to be given to the XMM-Newton helpdesk team and the SAS developers, for their time and patience helping address the many SAS-related problems with the development of the analysis pipeline. I would like to thank Dominique Eckert, Jelle de Plaa, and Jelle Kaastra, for several discussions and method-related motivation; Peter Boorman and Johannes Buchner for their help with using BXA; and Michal Zaja\v{c}ek for many helpful discussions. The scientific results reported in this article are based in part on data obtained from the Chandra Data Archive. This research has made use of software provided by the Chandra X-ray Center (CXC) in the application packages \textsc{Ciao} and \textsc{Sherpa}; as well as NASA's High Energy Astrophysics Software (HEASoft) packages \textsc{Xspec}. The scientific results reported in this article are also based in part on observations obtained with XMM-Newton, an ESA science mission with instruments and contributions directly funded by ESA Member States and NASA. JPB, NW, and TP acknowledge the financial support of the GA\v{C}R EXPRO grant No. 21-13491X. The material is based upon work supported by NASA under award number 80GSFC21M0002. K.U. acknowledges support from the National Science and Technology Council of Taiwan (grant NSTC 112-2112-M-001-027-MY3) and the Academia Sinica Investigator Award (grant AS-IA-112-M04). L.D.M. has been supported by the French government, through the UCA\textsuperscript{J.E.D.I.} Investments in the Future project managed by the National Research Agency (ANR) with the reference number ANR-15-IDEX-01.




\bibliographystyle{mnras}
\bibliography{bib} 


\newpage
\appendix

\section{Automatic Background Fitting}\label{sec:autobkg}
Generally, due to accumulating ionizing doses from various orbit cycles, the quantum efficiency of CCD detectors slowly degrades over time, resulting in a change in the shape of the underlying continuum due to an increase of noise in different detector channels. The shape of the \chandra{} background continuum has historically been fairly stable \citep{acis_shape}. The shape of the \xmm{} background continuum however, has been observed to vary between 8\% and 20\% between quiet periods and maximum \citep{pnshape}, while also being sensitive to the cleaning method associated with the cleaning and removal of background flares \citep{xmm_backgroundmodeling}. Additionally, the relative strengths of the different X-ray fluorescence line emission components also differ across the detector area due to the shielding and other nearby electronics.

For the various, relatively small, spectral extraction regions, the background model derived from the full field-of-view of each detector is not the most appropriate model. However, for each region on the detector, the global, underlying spectral model created from the description of the full field-of-view is a good first-order approximation for the start of the automatic background fitting procedure. 

The background contributions are scaled proportionally to the source and background region sizes in arcmin$^2$, by the exposure times between the source and backgrounds, and also by the high energy count rates in the 10 keV to 12 keV bands. Because the instrumental backgrounds originate from the detectors and are not dependent on the effective area of the optics, the instrumental background model should not be convolved by an Ancillary Response File (ARF); however, due to peculiarities with how models are set up in Sherpa, we instead build and convolve our model with a flat-ARF, which behaves the same as no ARF when convolved. There is a standing debate with regards to which is the correct response matrix to be used for the background since, arguably, a `correct' response would need to be calibrated inside the spacecraft during flight for the detector background specifically, and a separate additional response would need to be done for the soft proton (and other particle) contributions. We explored the fit results when using the photon-derived response matrix calculated from standard spectral extraction tasks and compared it with the fit results when using a diagonal-response matrix, which effectively provides infinite instrument resolution for the background, and found that the results were comparable within a few per cent. 

The background fitting procedure used in this paper works iteratively, carefully thawing and fitting model parameters in sequence, and becomes more fine-tuned as each model parameter is fit for the specific detector region. The background model is first normalized with a constant before fitting the continuum emission. The continuum is then fixed before the Gaussian line emissions are then subsequently thawn and fit, adjusting their shapes and relative ratios. Finally the normalization and continuum are then refit to slightly correct for the adjusted Gaussian lines.

The automatic background fitting routine initially defaults to fitting the full instrumental background first before fitting the source spectrum. By using robust instrumental background models, the typically difficult probes into faint signatures in poor S/N and S/b data regimes (such as cluster outskirts and filamentary structures close to $R_{200}$) can be exploited by leveraging the control over the systematics in the various backgrounds, preventing a loss of information through averaging or over smoothing of subtle spectral features which might be present in the data \citep{silvano_prospective_outskirts}. This background modelling procedure is additionally motivated by a well-known bias when subtracting backgrounds using `wstat' statistics that treats every bin independently from one another (See Notebook \footnote{https://giacomov.github.io/Bias-in-profile-poisson-likelihood/}). This `wstat' bias is strongly affected by how spectra are binned, especially in low-resolution data, e.g., \rosat{}. Modelling the background and using `cstat' decreases the overall fit uncertainties when compared to `wstat' because it treats all of the bins as continuous, even in the case of low or zero-count bins. 

The caveat of this background fitting method, in particular, is that the background fitting procedure begins to fail when reaching very small regions with too few background counts. In other words, when the X-ray surface brightness is high (high S/N), the regions become small, which also increases the S/b and results in poor statistics.
Coincidentally, this happens to be near the size of the PSF of each instrument. In these situations, the many parameters of the complex background model cannot be properly constrained. Therefore, the model becomes inappropriate with regard to the data quality of the background data. In these cases, we simplify the complex model to adjust to the data quality. We begin to do this by first removing the instrumental Gaussian lines so that we can fit a constant and the continuum. Then we try to also remove the continuum emission, and simply fit a constant model to the background. After each of these fits are performed, the reduced statistic, Akaike Information Criterion (AIC), and Bayes Information Criterion (BIC) are computed, and the most appropriate background model is selected for the dataset.

For the very extreme cases, other background modelling methods and tools exist which could also be used; however, in this high S/b regime, a debate could be made about the need for a background model in the first place since the background can effectively be treated as noise \citep{bxa, simmonds, suzuki21}. 

The AIC and the BIC both give us indicators for model selection based on the computed evidence \citep{akaike, akaike_corrected}. The small-sample size corrected AIC and BIC are described, respectively, via,
\begin{align}
    AIC_c &= 2k - 2\ln(\mathcal{L}) + \frac{2k^2 + 2k}{n - k -1}\label{eq:akaikec}  \,,\\
    BIC &= k \ln(n) - 2\ln(\mathcal{L})\,,
\end{align}
where $k$ is the degrees of freedom in the model, $n$ is the number of data points, and $\mathcal{L}$ is the maximized value of the likelihood function. The non-corrected AIC equation is the same as Eq.~\ref{eq:akaikec} without the final term. 

As evident by the two equations, the AIC and BIC both try to account for the goodness of fit between models, with a penalty given for the number of model parameters, where respectively, the AIC penalty is 2$k$, versus the BIC penalty of $\ln(n)k$. Generally speaking, this punishes model complexity to prevent favouring models which overfit the data. These values can then be used to inform model selection; however, this is only applicable when comparing nested models relative to the same fitted data. The `cstat' value is approximately equal to $-1/2 \ln(\mathcal{L})$ and has been shown to be a reliable estimator for model performance even with as few as 30 counts \citep{cash,fit_goodness_cstat}. We can then modify Eq.~\ref{eq:akaikec} to give metrics for our background model selection.

Regardless of which background is chosen, the novelty of this tool is that the forward model is built completely from first principles, is intuitive, and is easily extendable, so new missions like \xrism and \athena will be able to benefit from it as well. The background modelling tool is open source and will be released alongside a full pipeline for both \chandra{} and \xmm{}, which handles the intermediate processing of the spectra\footnote{https://github.com/jpbreuer}.

\section{Instrumental Cross-calibration}\label{sec:cross_calibration}
There have been many studies exploring the cross-calibration between \chandra{}, \xmm{}, and other X-ray observatories, motivated by the International Astronomical Consortium for High-Energy Calibration (IACHEC) using supernova remnants, quasars, and relaxed galaxy clusters \citep[see][Section 6.1.2 for a detailed overview]{supernova_crosscalibration}. 

Using fluxes as a proxy, \citet{clusters_crosscalibration} found that the fluxes between \chandra{} ACIS and \xmm{} pn had a large scatter; however, they were consistent within 2\% in the soft band (0.5 - 2.0 keV) and that ACIS was, on average, 11\% higher in the hard band (2.0 - 7.0 keV). \citet{mospncalib}, on the other hand, showed a difference in ACIS fluxes between 0-10\% in the soft band and a 0-5\% difference between ACIS and \xmm{} MOS. A more recent study discusses an agreement on flux between ACIS and MOS below 3 keV, but the flux discrepancy rises to 10-15\% above 4 keV \citep{iachec_crosscalibration}. 

Regarding measured temperatures, previous studies have shown a 5-10\% difference on the effective temperature between the MOS and pn detectors \citep{mospncalib}. \chandra{} systematically has shown higher temperatures than \xmm{} in the broad band, yet in another more recent study using \chandra{} and \nustar{} data, \citet{chandra_nustar_crosscal} showed a 10.5\% and a 15.7\% temperature discrepancy between the instruments using the broad band (0.6 - 9.0 keV) and hard band (3.0 - 10.0 keV), respectively. An even more recent cross-calibration talk from the 2023 IACHEC meeting\footnote{https://iachec.org/wp-content/presentations/2023/WikIACHEC2023.pdf} elaborated on this study further and discusses a complex relationship between the cross-calibrations between \chandra{} and \xmm{}, using \nustar{} as a reference. This talk mentions that the temperature discrepancies between instruments increase with respect to measured temperatures, also noting a $\sim$20\% flux excess for \xmm{} in the soft band. Furthermore, there is evidence that \xmm{} agrees better with \nustar{} for hot clusters, while \chandra{} agrees better for cooler clusters.

We performed several independent tests to explore the cross-calibrations between the various instruments as part of the verification of our joint fitting methodology using the instrumental background models. We found a consistent, 12-17\% cross-calibration related uncertainty between \chandra{} and \xmm{} between the three different observations of \macscluster{} and a 1-3\% uncertainty between the MOS and pn detectors. To do this, we fit a circular region from the high S/N regime of \macscluster{}, where there is an extremely hot ($\gtrsim 20$ keV) plasma emission, using a collisional ionisation equilibrium (\texttt{apec}) plasma model multiplied by a constant. We further explored the effect that the scaling of the instrumental backgrounds has on the fit results by scaling the \chandra{} backgrounds by the number of counts and also by scaling by the count-rate in the 10-12 keV band. We found that the temperatures were consistent between \chandra{} and \xmm{} when the background was scaled by the ratio of the observation to background high-energy particle count rates, but that the normalizations were off by around 18\%. However, when the background is instead scaled by the ratio of the high-energy number of counts rather than by exposure time, the normalizations were within 5\%, yet the temperatures were then off by around 22\%. Despite our measured discrepancies between the different instruments, we found that our uncertainties fell within the previously reported limits from different IACHEC studies, so our new instrumental background modelling approach does not seem to be introducing any new biases to the measurements that were not already previously known or reported. 

\begin{figure}[h!]
\centering
\includegraphics[width=\columnwidth]{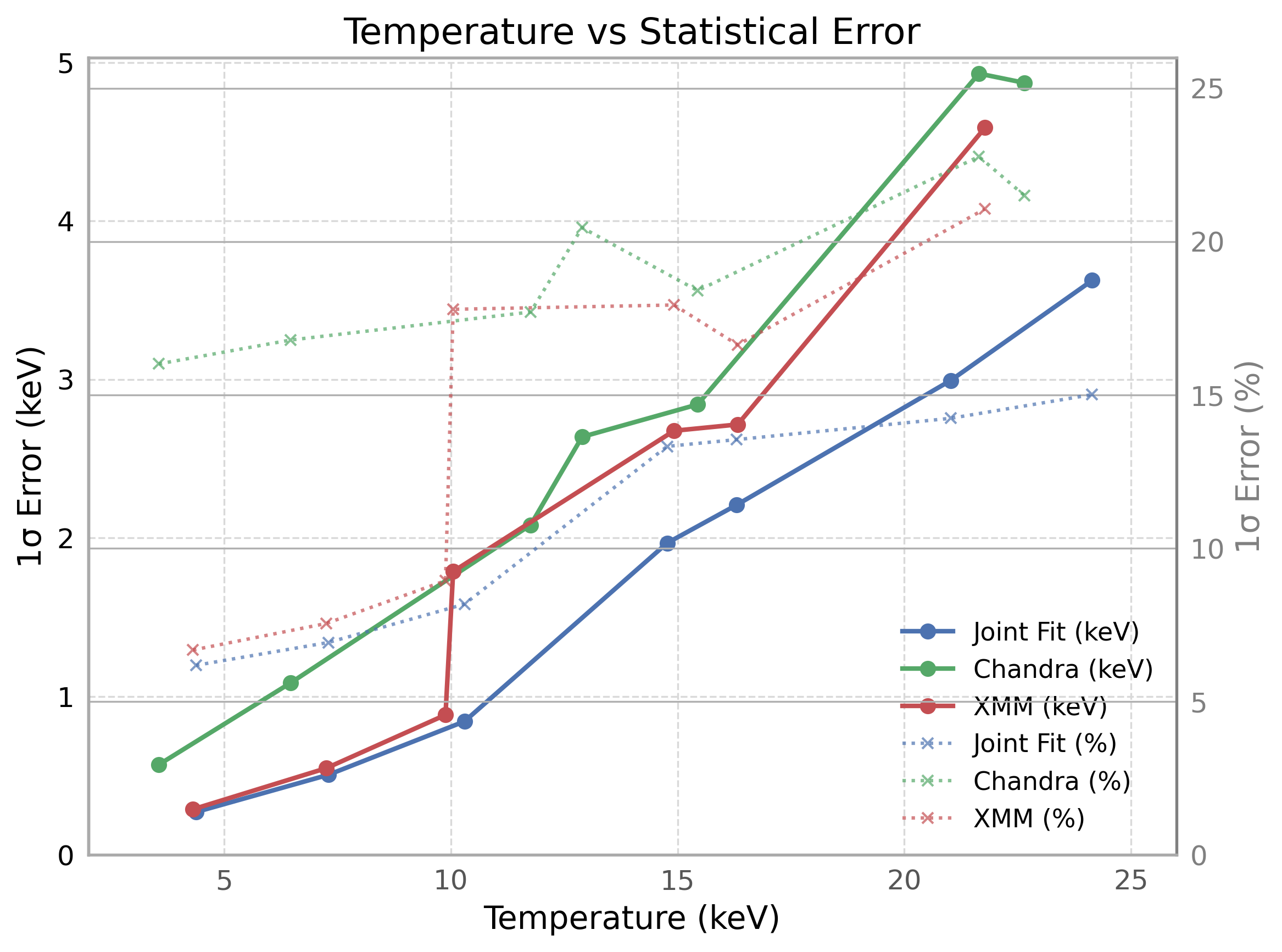}
\caption{Plot of the 1 sigma error as a function of temperature for \chandra{}, \xmm{}, and the combined fitting procedure in units of keV and as percents.\label{fig:syserr}}
\end{figure}

Fig.~\ref{fig:syserr} shows the statistical errors of the various instruments as a function of temperature using our joint fitting method, by extracting spectra from various regions with different temperatures. For comparison, the systematic uncertainties for temperatures of up to $\sim10$ keV, are estimated to be of the order of 10--15\% \citet{clusters_crosscalibration}, while for higher temperatures, based on the studies of the Bullet cluster, they are around 20\%  \citep{markevitch2006,wik2014}. 

\newpage
\section{\mustang{} Full FOV Image}
\label{sec:mustang_fov}

\begin{figure}[h!]
\centering
\includegraphics[width=\columnwidth]{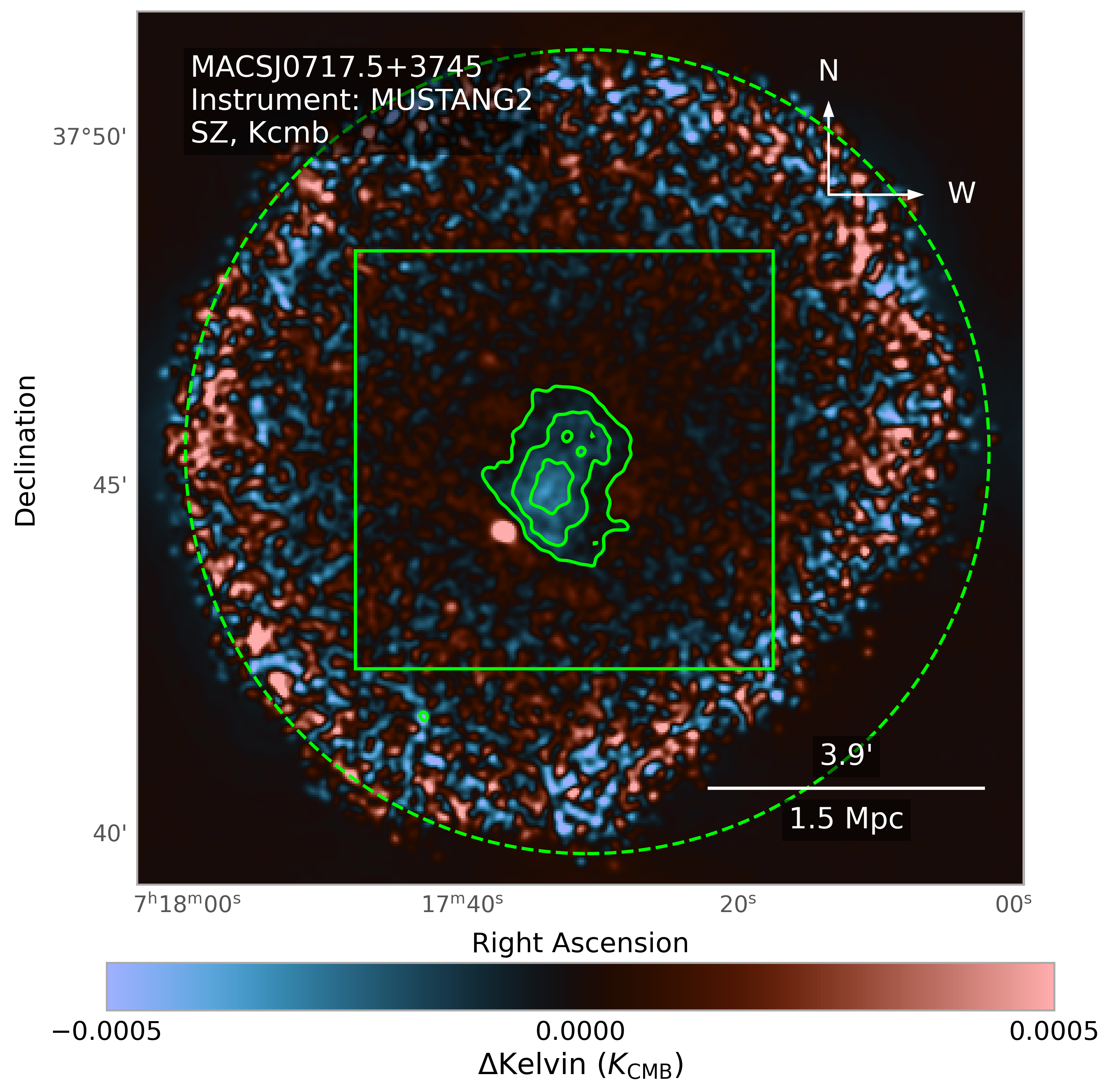}
\caption{FOV of the signal to noise ratio of the \mustang{} data shown in Fig.~\ref{fig:sz}. The FOV corresponds to the white box region shown throughout the paper, while the green box, and contours are the same as described in Fig.~\ref{fig:sz}.}\label{fig:m2fov}
\end{figure}

\begin{figure}[h!]
\centering
\includegraphics[width=\columnwidth]{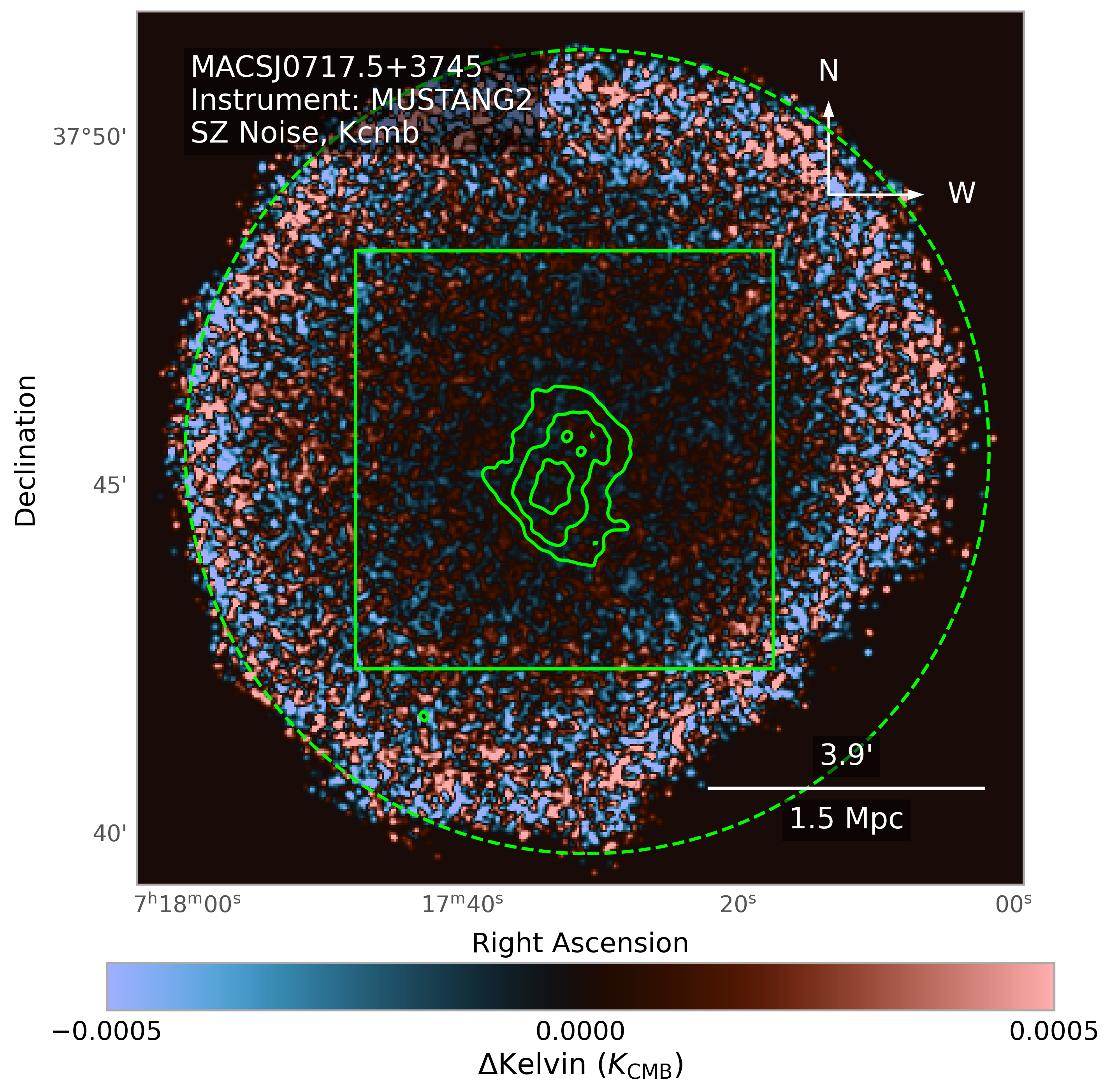}
\caption{A realization of the \mustang{} noise map across the FOV shown in Fig.~\ref{fig:m2fov}. The FOV corresponds to the white box region shown throughout the paper, while the green box, and contours are the same as described in Fig.~\ref{fig:sz}.}\label{fig:m2noise}
\end{figure}

\newpage
\section{3D Line of Sight Adjusted Maps and Projection Effects}\label{sec:los_estimate}

\begin{figure*}
\centering
\includegraphics[width=\textwidth]{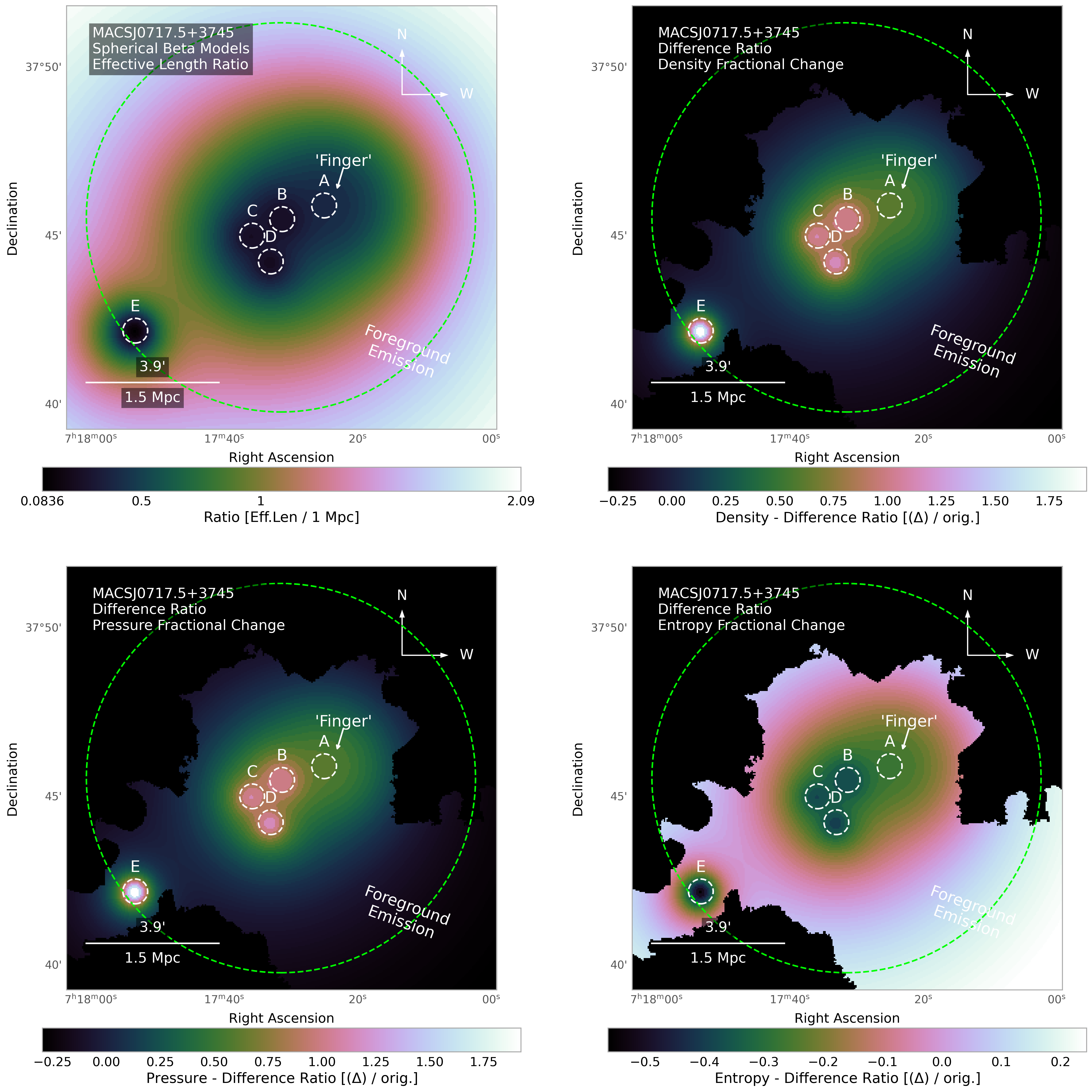}
\caption{Fractional differences between the corrected thermodynamic maps and the original maps, illustrating the scale of corrections due to the assumed LoS geometry.}\label{fig:frac_diff}
\end{figure*}

\begin{figure*}
\centering
\includegraphics[width=\textwidth]{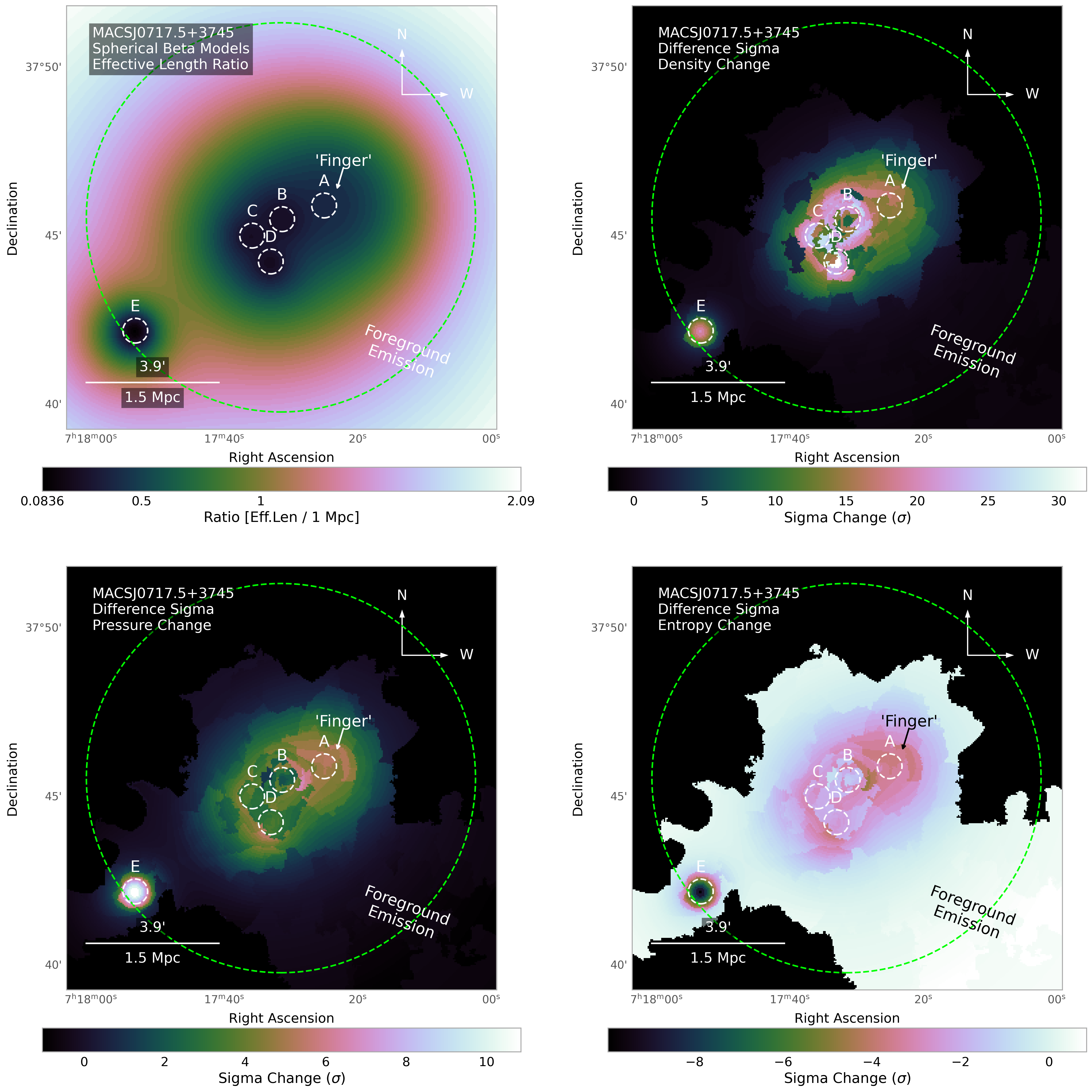}
\caption{Differences between the corrected and original maps expressed in units of sigma ($\sigma$), calculated from the statistical uncertainties associated with each bin.}\label{fig:sigma_diff}
\end{figure*}

The magnitude and spatial distribution of the geometric correction are clearly highlighted in Fig.\ref{fig:frac_diff}, showing fractional differences between corrected and original maps. Additionally, Fig.\ref{fig:sigma_diff} presents these differences expressed in terms of statistical significance ($\sigma$), based on the propagated uncertainties calculated from the original spectral fit for each independent bin. We caution the reader that, due to this being a highly disturbed, merging cluster with at least five subclusters in various stages of merger, the geometry assumed for this estimate is highly uncertain.

\label{lastpage}
\end{document}